\documentclass[acmsmall, screen]{acmart}
\usepackage{booktabs} 

\usepackage{hyperref}
\usepackage{ragged2e}
\usepackage{graphicx}
\usepackage{multirow}
\usepackage{graphicx}
\usepackage[update,prepend]{epstopdf}
\usepackage{balance}
\usepackage[linesnumbered, ruled]{algorithm2e}
\usepackage{algpseudocode}
\usepackage{caption}
\usepackage{subcaption}
\usepackage{footnote}
\usepackage[T1]{fontenc}
\usepackage[utf8]{inputenc}
\usepackage{tabularx,ragged2e,booktabs}
\usepackage{mathrsfs}
\usepackage{amsmath}

\usepackage{amssymb}
\usepackage{amsthm}
\usepackage{cases}
\usepackage{listings}
\usepackage{xcolor}
\usepackage{comment}
\usepackage{amsfonts}
\usepackage{yyy-math}
\usepackage{textcase}
\usepackage{wrapfig}
\usepackage{graphicx}
\usepackage{xcolor}
\usepackage{subcaption}
\usepackage{epstopdf}
\usepackage{url}
\usepackage{tabularx}
\usepackage{paralist}
\usepackage{amssymb}
\usepackage{pifont}

\newtheorem{example}{\textbf{Example}}

\newcolumntype{C}[1]{>{\Centering}m{#1}}

\DeclareMathOperator*{\argminyyy}{arg\,min}
\newcommand*{\argminl}{\argminyyy\limits}

\usepackage{color}
\definecolor{dkgreen}{rgb}{0,0.6,0}
\definecolor{gray}{rgb}{0.5,0.5,0.5}
\definecolor{mauve}{rgb}{0.58,0,0.82}
\newcommand{\oldsystem}{{\sc MatFast}}
\newcommand{\system}{{\sc MatRel}}

\setcopyright{acmcopyright}

\acmDOI{0000001.0000001}

\begin{document}
\title{Scalable Relational Query Processing on Big Matrix Data}

\author{Yongyang Yu}
\affiliation{%
  \institution{Facebook}
  \streetaddress{1 Hacker Way}
  \city{Menlo Park}
  \state{CA}
  \postcode{94025}
  \country{USA}}
\email{yongyangyu@fb.com}
\author{Mingjie Tang}
\affiliation{%
  \institution{Chinese Academy of Science}
  \streetaddress{Zhongguan district}
  \city{Beijing}
  \state{Beijing}
  \postcode{10010}
  \country{China}
}
\email{tangrock@gmail.com}
\author{Walid G. Aref}
\affiliation{%
 \institution{Purdue University}
 \streetaddress{305 N. University Street}
 \city{West Lafayette}
 \state{IN}
 \postcode{47907}
 \country{USA}}
\email{aref@purdue.edu}

\renewcommand\shortauthors{Yu, Y. et al}

\begin{abstract}
The use of large-scale machine learning methods is becoming ubiquitous in many applications ranging from business intelligence to self-driving cars. These methods require a complex computation pipeline consisting of various types of operations, e.g., relational operations for pre-processing or post-processing the dataset, and matrix operations for core model computations. Many existing systems focus on efficiently processing matrix-only operations, and assume 
that
the inputs to 
the 
relational operators are already pre-computed and are materialized as  intermediate matrices. However, the input to a relational operator may be complex in machine learning pipelines, and may involve various combinations of matrix operators. 
Hence, it is critical to realize scalable and efficient relational query processors that directly operate
on big matrix data.
This paper presents new efficient and scalable relational query processing techniques on big matrix data for in-memory distributed clusters. The proposed techniques leverage algebraic transformation rules to rewrite query execution plans into ones with lower computation costs. A distributed query plan optimizer exploits the sparsity-inducing property of merge functions as well as Bloom join strategies for efficiently evaluating various flavors of the join operation.
Furthermore, optimized partitioning schemes for the input matrices are developed to facilitate the performance of join operations based on a cost model that minimizes the communication overhead. The proposed relational query processing techniques are prototyped in Apache Spark. Experiments on both real and synthetic data demonstrate that the proposed techniques achieve up to two orders of magnitude in performance improvement over state-of-the-art systems on a wide range of applications.
\end{abstract}

%
%


%
%


\maketitle

\section{Introduction}
Data analytics, including machine learning (ML, for short) and scientific research, often need to analyze large volumes of matrix data in various applications, e.g., business intelligence~(BI), 
natural language processing~\cite{Manning:2003}, social network analysis, recommender systems~\cite{Beel:2016}, and genomic diagnostics~\cite{eqtl}. As ML models increase in complexity, and data volume accumulates drastically, traditional single-node solutions are incapable of processing this data efficiently. As a result, a number of distributed systems have been built to optimize data processing pipelines in a unified ecosystem of big matrix data for various kinds of applications.

In addition, queries to ML and scientific applications exhibit quite different characteristics from traditional business data processing applications. Complex analytics, e.g., linear regression~(LR, for short) for classification, and principle component analysis~(PCA) for dimension reduction, are prevalent in these query workloads.
These complex analytics replace the traditional SQL aggregations that are commonly used in BI applications. These analytical tasks rely  on large-scale matrix computations, and are more CPU-intensive than typical relational database computations.

The advent of MapReduce~\cite{mapreduce} and Apache Spark~\cite{Zaharia:spark} has spurred a number of distributed matrix computation systems, e.g., Mahout~\cite{mahout}, SystemML~\cite{icde11:systemml}, DMac~\cite{sigmod15:DMac}, MLlib~\cite{mllib:link}, and \oldsystem~\cite{matfast}. In contrast to popular scientific platforms, e.g., R~\cite{r-language}, the above systems provide better scalability and fault-tolerance in addition to efficiency in computations. 
However, they only focus on the efficient execution of \textit{pure} matrix computations, where the computation consists of \textit{only} matrix operations, e.g., matrix-matrix multiplications.

Data analytics on real-world applications need an entire data processing pipeline that consists of multiple stages and various data operations, e.g., relational selections and aggregations on matrix data. 
A typical pipeline of ML and scientific applications consists of several components, e.g., data pre-processing, core computation, and post-processing. 

\begin{figure}[!t]
\centering
\includegraphics[width=0.65\columnwidth]{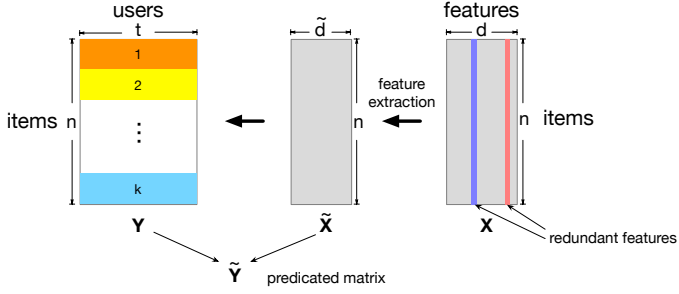}
\caption{Collaborative filtering with side information.}
\label{fig:rec_sys}
\vspace{-1.5em}
\end{figure}

\begin{example}[Collaborative Filtering]
Consider collaborative filtering with side information~\cite{side-info-cf}. In Figure~\ref{fig:rec_sys}, an $n \times t$ item-user recommendation matrix $\mY$ contains the observed recommendations of $t$ users over $n$ items. Entry $Y_{ij} \ne 0$ indicates that Item $i$ is recommended by User $j$. An $n \times d$ item-feature Matrix $\mX$ contains side information, where each row is a feature vector for the corresponding item. The goal is to identify the most-promising top-$k$ items for each target user from her non-recommended items based on $\mY$ and $\mX$. However, Matrix $\mX$ needs to be constructed, e.g., by crawling each item's webpage. Certain features may contain inconsistent or empty values. For successful feature extraction, all the empty columns are removed from $\mX$ and any inconsistent data is corrected by a data cleaning toolkit, i.e., matrix column and entry selection. Cross-validation~\cite{mlbook} is a common technique to estimate the regularizers to prevent over-fitting. The $k$-fold cross-validation divides the training dataset into $k$ disjoint partitions, selects $(k-1)$ partitions as the training set, and keeps the $k$-th partition for testing. 
The generation of $k$ disjoint partitions can be achieved by relational selections on the row dimension of Matrix $\mY$. 
For the post-processing, a max aggregation is conducted on the predicated matrix $\tilde{\mY}$ that is computed from the collaborative filtering model using $\mY \Join \tilde{\mX}$, where $\mY$ joins $\tilde{\mX}$ on the row dimension, to find the entries with the largest predicted values to recommend them to potential users. The matrix-only operations cannot evaluate the entire ML pipeline as certain relational operations cannot be computed, e.g., matrix-matrix join.
The example suggests that the successful training and prediction of a complex ML model requires a seamless execution of high-performance relational as well as linear-algebra operations.
\end{example}

Although several systems provide efficient matrix computations, they do not offer relational operations over matrix data, or only provide limited optimizations. SciDB~\cite{sigmod10:scidb} is an array database system that supports rich relational operations on array data. However, SciDB is not well-tuned for sparse matrix computations. 
For other systems built on general dataflow platforms, e.g., SystemML, MLlib, \oldsystem, the relational operations could become a bottleneck in the entire matrix data processing pipeline if they are not treated properly. An effective matrix query optimizer should be able to process both relational and matrix operations efficiently, and discover the potential to execute the mixed types of operations interchangeably, e.g., pushing a relational operation under a matrix operation in the logical query evaluation plan with correctness guarantees. 

\begin{figure}[!t]
\centering
\includegraphics[width=0.4\columnwidth]{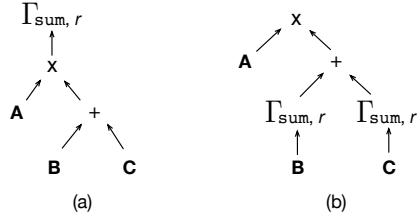}
\caption{Pushing aggregation under matrix multiplication.
$\Gamma_{\texttt{sum},r}$ is \texttt{sum} in the row direction. Essentially, Plan~(a) evaluates $\sum_j\sum_k a_{ik}*(b_{kj} + c_{kj})$ while  Plan~(b) evaluates $\sum_k a_{ik}*(\sum_j b_{kj} + \sum_j c_{kj})$.}
\vspace{-1em}
\label{fig:agg_over_mult}
\end{figure}

In this paper, we identify equivalent transformation rules for rewriting logical query plans that involve both relational and matrix operations. 
For example, a query optimizer that is not sensitive to matrix operations evaluates $\Gamma_{\texttt{sum},r}(\mA \times (\mB + \mC))$ as in Figure~\ref{fig:agg_over_mult}a.
The improved logical plan in Figure~\ref{fig:agg_over_mult}b executes the aggregate function before matrix multiplication. This plan is cheaper as the dimensions of the summed results are much smaller, hence reducing the cost of the matrix-matrix multiplication.
The query optimizer should be able to enumerate all the equivalent transformations and their corresponding plans, and pick the plan with the minimum computational cost.
Also, in a distributed setup, the query optimizer for plans with mixed relational and matrix operations should be aware of the communication overhead incurred by expensive relational operations over big matrices.
Load-balanced matrix data partitioning schemes, e.g., hash-based schemes, may not be efficient for joins with data dependencies. Consider the collaborative filtering example. The predicted item-user recommendation matrix $\tilde{\mY}$ could be computed from two factor-matrices $\mW$
and $\mH$ learned from the observed data $\mY$ and $\tilde{\mX}$, e.g., $\tilde{\mY} = \mW \times \mH^T$. Suppose that we have another user-item matrix $\mC$ with the same users as $\tilde{\mY}$ on different items. Tensor decomposition~\cite{Kolda:tensor} allows to capture connections among the users and two different sets of items. As will be explained in this paper, 
this tensor is constructed by conducting a join operation on two matrices, i.e., $\tilde{\mY} \Join \mC$, or $(\mW \times \mH^T) \Join \mC$, on the column dimension of $\tilde{\mY}$ and the row dimension of $\mC$. 
An efficient execution plan would partition Matrix $\tilde{\mY}$ in the column dimension and Matrix $\mC$ in the row dimension to satisfy the requirement of the join predicate. Partitioning Matrix $\tilde{\mY}$ in the column dimension further requires evaluating the costs of different matrix data partitioning strategies on $\mW$ and $\mH$, and choosing the one with the minimum communication overhead. Therefore, optimizing the data partitioning of the input and intermediate matrices is a critical step for efficiently evaluating relational operations over big matrix data.

In this paper, we introduce \system, an in-memory distributed relational query processing system on big matrix data. \system~has (1)~a query optimizer to identify and leverage transformation rules of the interleaved relational and matrix operations to reduce the computation cost and the memory footprint, and (2)~a matrix data partitioner to reduce the communication cost for relational joins over matrix data. 
The optimizer utilizes rule-based query rewrite to generate an optimized logical plan. 
\system~leverages a cost model to partition the input matrices for relational join operations to minimize communication overhead. \system~extends Spark SQL by providing a series of matrix operations, and leverages Spark SQL's relational query optimization techniques and dataflow operators.

The main contributions of this paper are as follows: 
\begin{itemize}
\setlength\itemsep{-1.2em}
\item We develop \system, an in-memory distributed system for relational query processing over big matrix data. \\
\item We identify equivalence transformation rules to rewrite a logical plan involving both relational and matrix operations. \\
\item We define join operators over matrix data and introduce optimization techniques to enhance their runtime cost. \\ 
\item We introduce a cost model to distribute matrix data for communication-efficient relational join operations.\\
\item We evaluate \system~against state-of-the-art distributed matrix computation systems using real and synthetic datasets. Experimental results illustrate \system's up to two orders of magnitude enhancement in performance.
\end{itemize}

The rest of this paper proceeds as follows.
Section~\ref{sec:architecture} presents 
the supported matrix operations and \system's architecture. 
Sections~\ref{sec:rel-algebra} and~\ref{sec:join} present the formal definitions of the supported relational operations and their corresponding optimizations over matrix data. 
Section~\ref{sec:impl} discusses
the local execution of operators and the
system implementation. 
Section~\ref{sec:experiment} presents experimental results. 
Section~\ref{sec:related-work} 
presents the related work, and Section~\ref{sec:conclusion} concludes the paper.

\section{Preliminaries}
\label{sec:architecture}

{\bf Notation:}\\
We adopt the following conventions: (1)~A bold, upper-case Roman letter, e.g., $\mA$, denotes a matrix. 
(2)~A regular Roman letter with subscripts, e.g., $A_{ij}$, is an element in Matrix $A$. 
(3)~A bold, lower-case Roman letter, e.g., $\vx$  is a column vector. 
(4)~A lower-case Greek letter, e.g.,  $\beta$, is a scalar. (5)~$\mA_{ij}$ is a block matrix, where $i$ and $j$ are the row- and column-block indexes.  
(6)~$(X_{ij})$ is a matrix with Element $X_{ij}$ at the $i$-th row and the $j$-th column. 
(7)~A bold, upper-case calligraphic Roman letter, e.g., $\tensor A$, is a tensor.
(8)~A regular calligraphic Roman letter with subscripts, e.g., $\elm A_{ijk}$ or $\elm A_{ijk\ell}$, is an element in a 3{rd}- or 4{th}-order tensor~\cite{Kolda:tensor}. The \textit{order} of a tensor is the number of dimensions.

\noindent
{\bf Supported Matrix Operators}:\\
\system~extends \oldsystem~\cite{matfast}. \system~supports the unary matrix transpose operator $\mB = \mA^T, B_{ij} = A_{ji}, \forall i,j$; following binary matrix operators: matrix-scalar addition, $\mB = \mA + \beta, B_{ij} = A_{ij} + \beta$; matrix-scalar multiplication, $\mB = \mA * \beta, B_{ij} = A_{ij} * \beta$; matrix-matrix addition, matrix-matrix element-wise multiplication/division, $\mC = \mA \star \mB, C_{ij} = A_{ij} \star B_{ij}$, where $\star \in \{+, *, /\}$; and matrix-matrix multiplication, $\mC = \mA \times \mB, C_{ij} = \sum_k A_{ik}*B_{kj}$. By default, matrix transpose has the highest precedence, and matrix-matrix multiplication has a higher precedence over element-wise operators. Besides basic matrix operators, many advanced operations are also supported in terms of basic ones~\cite{matrixBook}, e.g., matrix inverse and \textit{QR} factorization. In addition, \system~supports relational operations, e.g., selection, aggregation, and join. 

\noindent
{\bf Overview of \system}: \\
\begin{figure}
\centering
\includegraphics[width=0.6\columnwidth]{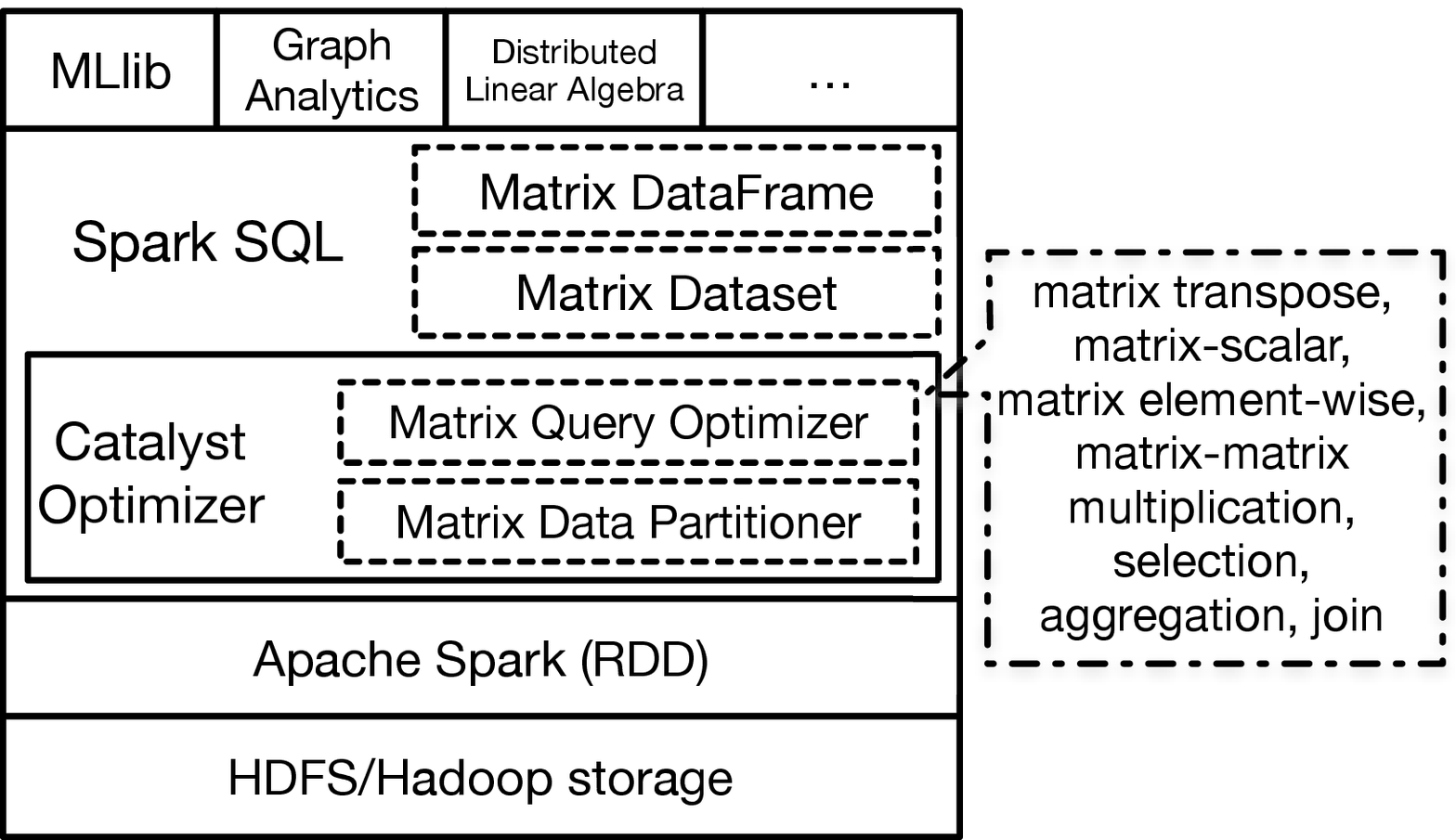}
\caption{Architecture of \system.}
\vspace{-1em}
\label{fig:overview}
\end{figure}
Refer to Figure~\ref{fig:overview}. 
The input to \system~can be an ML algorithm or a graph analytical task that consists of various relational and matrix operators over the input matrices. 
\system~organizes the operators into a logical evaluation plan that is fed to \system's query optimizer.
\system's optimizer extends Spark SQL's Catalyst optimizer~\cite{sparksql} by introducing new relational data types for matrix data. 
The optimizer casts matrix data as relational tables with predefined schemas. 
Using rule-based query rewrite heuristics, the initial query plan is transformed into an optimized execution plan in a distributed environment. The optimizer invokes the matrix data partitioner to assign efficient partitioning schemes to the distributed matrix data based on a cost model. Internally, the matrix data is organized as Resilient Distributed Datasets~(RDDs)~\cite{Zaharia:spark} for distributed execution. For efficient and fault-tolerant storage of matrix data, \system~uses Hadoop Distributed File System~(HDFS) for external storage.

\section{Relational operators on matrix data}
\label{sec:rel-algebra}


\noindent{\bf Relational Algebra for Matrix Data}: \\
Relational algebra~\cite{Codd:1970} and its primitive operators operate over relations. All relational operators accept, as input, relations with properly defined schemas, and produce, as output, relations with certain schemas that are derived from the input schemas. 
To ensure expressive relational operators over matrix data, \system~ casts a general schema over all matrix data. 

\subsection{Relational Schema of a Matrix}
\label{section:schema}
A matrix, say $\mA$, consists of two dimensional entries, $A_{ij}$. 
Each entry in a matrix resembles a tuple in a table with schema: \texttt{matrixA (RID, CID, val)},
where \texttt{RID} and \texttt{CID} are the row and column dimensions, respectively, and \texttt{val} is the value of the matrix entry.
In addition, \system~ maintains the dimensions of the matrix in the system catalog, i.e., the number of rows and columns.
This schema specifies the logical structure of a matrix. However, it does \textit{not} reflect how the matrix is physically stored. For efficiency and storage considerations, a matrix is divided into ``blocks'', i.e., it is partitioned into smaller matrices that can be moved to specific compute nodes for high-performance local computations. Detailed physical storage of matrices is presented in Section~\ref{physical-storage-matrix}.

\subsection{Relational Selection on Matrix Data}
Relational select~($\sigma$) selects matrix entries that qualify a certain predicate. 
We distinguish between selections on the dimensions and selections on the values of a matrix. 
A relational select is a unary operator of the form: 
$
\sigma_{\theta(RID, CID, val)}(\mA),
$
where $\theta$ is a propositional formula on the dimensions and entries of Matrix $\mA$. An atom in $\theta$ is defined as $u \varphi c$ or $u \varphi v$, where $u, v \in \{RID, CID, val\}$, $u \neq v$, and $c$ is a constant, $\varphi \in \{<, \le, =, \neq, \ge, >\}$; and the logical expressions among atoms, i.e., $\land$~(and), $\lor$~(or), and $\neg$~(negation) expressions. A relational select on a matrix produces another matrix, perhaps with different dimensions. The dimensions are determined by the predicate, e.g.,  
the 
predicate $``RID=i"$ produces a matrix of a single row while the predicate $``RID = CID"$ produces the diagonal vector of the matrix.

\noindent{\bf Selecting the Entries}:\\
A select produces a matrix of the same dimension if the select predicate only involves Attribute $val$. 
The matrix entries that qualify the select predicate are preserved, and the remaining entries are set to NULL.
The relational select on matrix entries does \emph{not} change the dimension of the input.  
If a series of select predicates apply over a matrix, the following transformation rule applies:
\begin{equation} \label{rule:1}
\sigma_{\theta_1}(\sigma_{\theta_2}(\cdots(\sigma_{\theta_{k}}(\mA)))) = \sigma_{\theta_1 \land \theta_2 \land \cdots \land \theta_k}(\mA),
\end{equation}
where each $\theta_i$ is a predicate on the matrix entries.

\noindent
{\bf Relational Select on the Matrix Dimensions}: \\
A select predicate can operate on the matrix dimensions to pick slices of a matrix according to the given row/column dimension value. 
For example, cross-validation~\cite{mlbook} is an ML technique to estimate how accurately a predicative model will perform in practice. Specifically, leave-\emph{one}-out cross-validation can be realized by selecting a single row from a training matrix as the test set. 
The output of a select on the dimensions is a matrix with different dimensions than the input's, e.g., if the input is an $m\times n$ Matrix $\mA$, $\sigma_{RID = i}(\mA)$ is a $1\times n$ matrix whose entries are the $i$-th row of $\mA$.
Similarly, $\sigma_{CID = j}(\mA)$ is an $m\times 1$ matrix 
whose entries are the $j$-th column; 
$\sigma_{RID \ge i_1 \land RID \le i_2}(\mA)$ produces a matrix slice of $\mA$, whose dimension is $(i_2 - i_1 + 1)$-by-$n$. 


The select predicate can be composed of conditions on both dimensions and entries.
For instance, selecting all the entries whose values equal 10 from the 5-th row is expressed by $\sigma_{val = 10 \land RID = 5}(\mA)$.
Furthermore, in real-world ML applications, an empty column suggests no metrics have been observed for some feature. Removing empty columns will benefit the feature extraction procedure for better efficiency. Thus, we introduce two special predicates,
$
\sigma_{rows \neq NULL}(\mA)~\text{and}~
\sigma_{cols \neq NULL}(\mA).
$
The former excludes the rows that are all NULLs, and the latter excludes the columns that are all NULLs. These are important extensions to the select operator because empty rows/columns usually do not contribute to matrix computations. 

\noindent
{\bf Optimizations for Selects over Dimensions}:\\
\system~supports matrix operations including transpose, matrix-scalar operation, matrix element-wise operation, and matrix-matrix multiplications. For matrix transpose, the transformation rules: $\sigma_{RID=i}(\mA^T) = (\sigma_{CID=i}(\mA))^T$ and $\sigma_{CID=j}(\mA^T) = (\sigma_{RID=j}(\mA))^T$ 
reduce the computational costs significantly by avoiding the matrix transpose. It avoids $O(n^2)$ operations to compute the matrix transpose by leveraging a select on a single dimension.

Similar rewrite rules apply to matrix-scalar and matrix element-wise operations. The following rules can be viewed as pushing the select below the matrix operations:
$\sigma_{RID=i}(\mA \star \mB) = \sigma_{RID=i}(\mA) \star \sigma_{RID=i}(\mB)$, 
where $\star \in \{+, *, /\}$. This rule
also applies to matrix-scalar multiplications. We illustrate these transformation rules on the selects w.r.t. the row dimension. The rules apply to selections on the column dimension as well. The number of required operations drops from $O(n^2)$ to $O(n)$ by circumventing the intermediate matrix.
For matrix-matrix multiplications, we have $\sigma_{RID=i}(\mA \times \mB) = \sigma_{RID=i}(\mA) \times \mB$.
\begin{proof}
Let $\mC = \mA \times \mB$, where $C_{ij} = \sum_k A_{ik} * B_{kj}$. Let $C_{i:}$ denote the $i$-th row of Matrix $\mC$. Then, by definition, $C_{i:} = \sum_k A_{ik} * B_{k:}$, where $A_{ik}$ is the $(i, k)$-th entry from $\mA$, and $B_{k:}$ is $k$-th row from $\mB$. This can be viewed as the summation of each entry in the $i$-th row of $\mA$ multiplying with corresponding rows of $\mB$, i.e., $\mC_{i:} = \mA_{i:} \times \mB$.
\end{proof}
Analogously, the Rule: $\sigma_{CID=j}(\mA \times \mB) = \mA \times \sigma_{CID=j}(\mB)$
can be verified. It takes $O(n^3)$ operations for matrix-matrix multiplications, while a matrix-vector multiplication costs $O(n^2)$ operations. 
Furthermore, the rule:
$\sigma_{RID=i \land CID=j}(\mA \times \mB) = \sigma_{RID=i}(\mA) \times \sigma_{CID=j}(\mB)$
reduces the complexity from $O(n^3)$ (matrix-matrix multiplication) to $O(n)$ (vector inner-product).

\subsection{Aggregation on Matrix Data}
Traditionally, an aggregation is  defined on the columns of a relation with various aggregate functions, e.g., \texttt{sum} and \texttt{count}. The aggregation operator is also beneficial on matrix data. For example, assume that for a user-movie rating database,
each row in the matrix represents a user, and each column represents a certain movie. Then, computing the average ratings of all the users can take place in the following way. 
The $(i,j)$-th entry in the user-movie rating matrix denotes the rating of $user_i$ on $movie_j$. This requires an average computation on the row dimension. Unlike relational data, aggregations on a matrix may occur in different dimensions. Formally, an aggregation is a unary operator, defined as
\[
\Gamma_{\rho,dim}(\mA),
\]
where $\rho$ is the name of the aggregate function, and $dim$ indicates the chosen dimension, i.e., $dim$ could be \emph{row}~(\emph{r}), \emph{column}~(\emph{c}), \emph{diagonal}~(\emph{d}), or \emph{all}~({\emph{a}}).
Specially, \system~supports various aggregate functions, $\rho \in \{\texttt{sum}, \texttt{nnz}$, $\texttt{avg}, \texttt{max}, \texttt{min}\}$. We introduce each aggregate function and corresponding optimizations in this section.

\noindent {\bf \texttt{Sum()} Aggregate:}\\
 \system~supports four different variants of \texttt{sum} aggregations $\Gamma_{\texttt{sum}, r/c/d/a}(\mA)$. 
We first introduce their definitions, then discuss optimization strategies. Given an $m$-by-$n$ Matrix $\mA$, $\Gamma_{\texttt{sum}, r}(\mA)$ produces an $m$-by-1 matrix, or a column vector, whose $i$-th entry computes the summation along the $i$-th row of $\mA$. Similarly,
$\Gamma_{\texttt{sum}, c}(\mA)$ generates a 1-by-$n$ matrix, or a row vector that computes the summations of entries along the column direction. $\Gamma_{\texttt{sum}, d}(\mA)$ is defined only for square matrices that is also called the \emph{trace} of a square matrix. The trace is simply a scalar value. Finally, $\Gamma_{\texttt{sum},a}(\mA)$ computes the summation of all the entries.

We present transformation rules for optimizing  \texttt{sum} aggregations when the input consists of various matrix operations. If the input is a matrix transpose, then we have:
\begin{align}
\Gamma_{\texttt{sum}, r/c}(\mA^T) &= (\Gamma_{\texttt{sum}, c/r}(\mA))^T, \label{rule:11}\\
\Gamma_{\texttt{sum}, d/a}(\mA^T) &= \Gamma_{\texttt{sum}, d/a}(\mA), \label{rule:12}
\end{align}
By pushing a \texttt{sum} aggregation before a matrix transpose, \system~computes the transpose on a matrix of smaller dimensions. For trace and all-entry summation, the evaluation of the transpose can be avoided.

For matrix-scalar operations, we can derive the following transformation rules,
\begin{align}
\Gamma_{\texttt{sum}, r/c/a}(\mA + \beta) &= \Gamma_{\texttt{sum}, r/c/a}(\mA) + \beta \ve_n/\beta \ve^T_m/\beta mn, \label{rule:13}\\
\Gamma_{\texttt{sum}, d}(\mA + \beta) &= \Gamma_{\texttt{sum}, d}(\mA) + \beta n, \label{rule:15}\\
\Gamma_{\texttt{sum}, \cdot}(\mA * \beta) &= \Gamma_{\texttt{sum}, \cdot}(\mA) * \beta, \label{rule:17}
\end{align}
where we denote $\ve_n$ to be an all-one column vector of length $n$ and Rule~\ref{rule:15} only holds for square matrices, i.e., $m=n$. For matrix-scalar multiplications, similar rules apply when the aggregation is executed along the column direction, the diagonal direction, and the entire matrix.

For matrix element-wise operations, only the matrix element-wise addition has a compatible transformation rule for \texttt{sum} aggregations.
\begin{align}
\Gamma_{\texttt{sum}, \cdot}(\mA + \mB) = \Gamma_{\texttt{sum}, \cdot}(\mA) + \Gamma_{\texttt{sum}, \cdot}(\mB). \label{rule:18}
\end{align}
Though Rule~\ref{rule:18} does not change the computation overhead, it circumvents storing the intermediate sum matrix $\mA+\mB$, alleviating memory footprint.

For matrix-matrix multiplications, we derive efficient transformation rules to compute the \texttt{sum} aggregations,
\begin{align}
\Gamma_{\texttt{sum}, r}(\mA \times \mB) &= \mA \times \Gamma_{\texttt{sum}, r}(\mB), \label{rule:19}\\
\Gamma_{\texttt{sum}, c}(\mA \times \mB) &= \Gamma_{\texttt{sum}, c}(\mA) \times \mB, \label{rule:20}\\
\Gamma_{\texttt{sum}, a}(\mA \times \mB) &= \Gamma_{\texttt{sum}, c}(\mA) \times \Gamma_{\texttt{sum}, r}(\mB), \label{rule:21}\\
\Gamma_{\texttt{sum}, d}(\mA \times \mB) &= \Gamma_{\texttt{sum}, a}(\mA^T * \mB) = \Gamma_{\texttt{sum}, a}(\mA * \mB^T). \label{rule:22}
\end{align}
\begin{proof}(Rule~\ref{rule:19})
Let $\mC = \mA \times \mB$, where $C_{ij} = \sum_{k}A_{ik} * B_{kj}$. Let the column Vector $\vy = \Gamma_{\texttt{sum}, r}(\mC)$. Therefore,
\[
y_i = \sum_j C_{ij}
=\sum_j \sum_k A_{ik} * B_{kj}
= \sum_k A_{ik} * b_k,
\]
where Vector $\vb = \Gamma_{\texttt{sum}, r}(\mB)$.
It is clear that the derivation in the matrix format, $\vy = \mA \times \vb$ that is \emph{exactly} Rule~\ref{rule:19}.
\end{proof}
Similar derivations also apply to Rules~\ref{rule:20} and~\ref{rule:21}. For Rule~\ref{rule:22}, both input matrices are required to be square. 
\begin{proof}(Rule~\ref{rule:22})
Let us write Matrix $\mA$ in terms of rows, and Matrix $\mB$ in terms of columns, i.e.,
\[
\mA =
\begin{bmatrix}
\va_1^T \\ \vdots \\ \va_n^T
\end{bmatrix}, \quad
\mB = 
\begin{bmatrix}
\vb_1 &\cdots &\vb_n
\end{bmatrix}.
\]
Therefore, the matrix-matrix multiplication can be represented by the vectors $\va_i$'s and $\vb_i$'s.
\[
\mA \times \mB = 
\begin{bmatrix} 
    \va_1^T\vb_1 & \dots & \va_1^T\vb_n \\
    \vdots & \ddots & \vdots \\
    \va_n^T\vb_1 & \dots & \va_n^T\vb_n 
\end{bmatrix}
\]
The trace of $\mA \times \mB$ is  computed as $\Gamma_{\texttt{sum}, d}(\mA \times \mB) = \sum_{i=1}^n \va_i^T \times \vb_i$.
The transpose of $\mA$ is denoted as,
\[
\mA^T = 
\begin{bmatrix}
\va_1 &\cdots &\va_n
\end{bmatrix},
\]
and the matrix element-wise multiplication is computed as:
\[
\mA^T * \mB = 
\begin{bmatrix}
\va_1*\vb_1 &\cdots &\va_n * \vb_n
\end{bmatrix}.
\]
When conducting the all-entry \texttt{sum} aggregation, it is essential to compute the inner-product of corresponding vector pairs, i.e., $\Gamma_{\texttt{sum},a}(\mA^T * \mB) = \sum_{i=1}^n \va_i^T \times \vb_i$. The second half of Rule~\ref{rule:22} can be verified in a similar manner.
\end{proof}
The computation complexity is usually $O(n^3)$ for square matrix-matrix multiplications. By using matrix element-wise multiplication and matrix transpose, the complexity is reduced to $O(n^2)$.

Unfortunately, equivalent transformation rules do not apply when the input involves a relational selection. 
For general selection condition $\theta$,
\begin{align}
\Gamma_{\texttt{sum}, \cdot}(\sigma_\theta(\mA)) &\neq \sigma_\theta(\Gamma_{\texttt{sum}, \cdot}(\mA)). \label{rule:23}
\end{align}
However, there exist valid transformation rules if the \texttt{sum} dimension happens to match with the predicate of the selection, e.g., $\Gamma_{\texttt{sum}, r}(\sigma_{RID=i}(\mA)) = \sigma_{RID=i}(\Gamma_{\texttt{sum}, r}(\mA))$. Swapping the execution order of \texttt{sum} aggregation leads to the same result. Pushing a select relational operation below a matrix operation is beneficial as it could reduce the dimension of the matrix.

\noindent {\bf \texttt{Count()} Aggregate:}\\
For relational data, the \texttt{count} aggregate function computes the number of tuples that satisfy a certain criteria. For matrix data, the \texttt{count} function is usually interpreted as computing the number of nonzeros~(\texttt{nnz}). This function is quite useful for optimizing matrix computations, e.g., sparsity estimation for sparse matrix chain multiplications~\cite{matfast}.
Similar to \texttt{sum} aggregation, \system~supports four \texttt{count} variants, $\Gamma_{\texttt{nnz},r/c/d/a}(\mA)$. 
The output of each function is defined as its counterpart of  the \texttt{sum} function correspondingly.

We present equivalent transformation rules for optimizing \texttt{count} aggregations when the input is composed of various matrix operations. For matrix transpose, we have:
\begin{align}
\Gamma_{\texttt{nnz}, r/c}(\mA^T) &= (\Gamma_{\texttt{nnz}, c/r}(\mA))^T, \label{rule:25}\\
\Gamma_{\texttt{nnz}, d/a}(\mA^T) &= \Gamma_{\texttt{nnz}, d/a}(\mA). \label{rule:27}
\end{align}
These rules push a \texttt{count} aggregation below a matrix transpose. The postponed matrix transpose is executed on a smaller matrix, e.g., a vector, avoiding $O(n^2)$ computations for the matrix transpose.

For matrix-scalar operations, let $\mA$ be an $m$-by-$n$ matrix, and scalar $\beta \neq 0$. The following  rules hold for an input that involves matrix-scalar operations,
\begin{align}
\Gamma_{\texttt{nnz}, r}(\mA + \beta) &= \ve_m * n, \label{rule:27}\\
\Gamma_{\texttt{nnz}, c}(\mA + \beta) &= \ve_n^T * m, \label{rule:28}\\
\Gamma_{\texttt{nnz}, d}(\mA + \beta) &= n, \label{rule:29}\\
\Gamma_{\texttt{nnz}, a}(\mA + \beta) &= mn, \label{rule:30}\\
\Gamma_{\texttt{nnz}, \cdot}(\mA * \beta) &= \Gamma_{\texttt{nnz}, \cdot}(\mA). \label{rule:31}
\end{align}
By assuming the constant $\beta \neq 0$, the number of zeros is not altered for matrix-scalar multiplications. For the matrix-scalar addition, the output does not depend on the entries in $\mA$, which only requires the dimensions to conduct the computation.

For matrix element-wise operations, only the division operator works properly with the \texttt{count} aggregation.
\begin{align}
\Gamma_{\texttt{nnz}, \cdot}(\mA / \mB) = \Gamma_{\texttt{nnz}, \cdot}(\mA) \label{rule:32}
\end{align}
Rule~\ref{rule:32} indicates that the matrix element-wise division preserves the number of nonzeros of the numerator matrix. Unfortunately, such convenient transformation rules do not apply to other matrix element-wise operators and matrix-matrix multiplications.

For an input consisting of relational selections, Rule~\ref{rule:23} can be extended too. It is impossible to swap the execution order of a \texttt{count} aggregation and a selection for a correct output. It is only valid to swap the execution order of a \texttt{count} aggregation and a selection when both work on the same dimension, e.g., $\Gamma_{\texttt{nnz}, c}(\sigma_{CID=j}(\mA)) = \sigma_{CID=j}(\Gamma_{\texttt{nnz}, c}(\mA))$.

\noindent {\bf \texttt{Avg()} Aggregate:}\\
The \texttt{avg} aggregate function computes the average statistic on a matrix in a certain dimension. \system~supports four different variants, $\Gamma_{\texttt{avg}, r/c/d/a}(\mA)$. 
The output of each \texttt{avg} aggregate function is defined as its counterpart of the \texttt{sum} aggregation correspondingly. The \texttt{avg} aggregation can be viewed as a compound operator from a \texttt{sum} and a \texttt{count} aggregations. Formally, any \texttt{avg} aggregation can be computed as:
\[
\Gamma_{\texttt{avg}, \cdot}(\mA) = \Gamma_{\texttt{sum},\cdot}(\mA) / \Gamma_{\texttt{nnz}, \cdot}(\mA).
\]
Therefore, optimizing an \texttt{avg} aggregation can leverage all the transformation rules we have explored for \texttt{sum} and \texttt{count} aggregations.
\system's optimizer automatically invokes query planning for 
\texttt{sum} and \texttt{count} aggregations when generating an execution plan for an \texttt{avg} aggregation.


\noindent {\bf \texttt{Max()}/\texttt{Min()} Aggregate:}\\
The \texttt{max}~(or \texttt{min}) aggregations are used to obtain the extreme values from a matrix, and serves as an important building block for anomaly detection~\cite{anomaly-detection-survey} in data mining applications. Naturally, \system~supports four \texttt{max}~(or \texttt{min}) variants, $\Gamma_{\texttt{max}/\texttt{min}, r/c/d/a}(\mA)$. 
The output of each aggregation is defined as its counterpart of the \texttt{sum} aggregation correspondingly.
The transformation rules for optimizing the \texttt{max}~(or \texttt{min}) aggregations can be derived as follows:
\begin{align}
\Gamma_{\texttt{max}/\texttt{min}, r/c}(\mA^T) &= (\Gamma_{\texttt{max}/\texttt{min}, c/r}(\mA))^T, \label{rule:33}\\
\Gamma_{\texttt{max}/\texttt{min}, d/a}(\mA^T) &= \Gamma_{\texttt{max}/\texttt{min}, d/a}(\mA), \label{rule:34} \\
\Gamma_{\texttt{max}/\texttt{min}, \cdot}(\mA + \beta) &= \Gamma_{\texttt{max}/\texttt{min}, \cdot}(\mA) + \beta, \label{rule:35}\\
\Gamma_{\texttt{max}/\texttt{min}, \cdot}(\mA * \beta) &= \Gamma_{\texttt{max}/\texttt{min}, \cdot}(\mA) * \beta \quad (\beta > 0), \label{rule:36}\\
\Gamma_{\texttt{max}/\texttt{min}, \cdot}(\mA * \beta) &= \Gamma_{\texttt{min}/\texttt{max}, \cdot}(\mA) * \beta \quad (\beta < 0). \label{rule:37}
\end{align}
We assume that the constant scalar $\beta \neq 0$. When $\beta < 0$, the optimizer computes the \texttt{min}/\texttt{max} aggregation instead 
of
avoiding the intermediate matrix $\mA * \beta$. For matrix element-wise operations and matrix-matrix multiplications, \textit{no} general rules apply.
For example, $\Gamma_{\texttt{min}, r}([1,1,2] + [3,0.5,0]) =\Gamma_{\texttt{min}, r}([4,1.5,2]) =1.5$, which could only be evaluated after computing the sum of the two matrices.
For an input consisting of relational selections, it is only valid to swap the execution order of a \texttt{max}/\texttt{min} aggregation and selection when both work on the same dimensions.




\section{Join on matrix data}
\label{sec:join}
The relational join~($\Join$) is a useful operator for picking corresponding entries that satisfy the join predicates from two separate matrices. For example, raster data overlay analysis~\cite{hadoopviz} requires a join on two matrices with matching row/column index, where a raster map can be interpreted as a matrix. 
Formally, a relational join is a binary operator,
\[
\mA \Join_{\gamma, f} \mB,
\]
where $\gamma$ is the join predicate, $f$ is the user-defined merge function that takes two matching entries and outputs a merging result, e.g., $z=f(x,y)$. We introduce various formats of $\gamma$ in different semantics in the subsequent sections. In general, the result of a relational join on two matrices is a \emph{tensor}, or a multi-dimensional array, whose dimensions are determined by the predicate $\gamma$.
We first discuss how to infer the schema of a join output based on the join predicates. Next, we introduce all the valid formats of $\gamma$. Finally, we investigate the optimization strategies for join executions.

\subsection{Schema of a Join Result}
Just like every matrix has a schema, 
\system~automatically produces a schema for a join result. The schema of a join result consists of two parts: the index and the value. In this work, we only consider equality predicates as the join condition. Given two input matrices $A (RID_A, CID_A, val_A)$ and $B (RID_B, CID_B, val_B)$, the cardinality of the dimension of $\mA \Join_{\gamma, f} \mB$ is 
\[
d = 4 - \delta_{dim},
\]
where $\delta_{dim}$ is the number of equality predicates on the join dimensions. For example, if the join predicates $\gamma=``RID_A=RID_B"$, the join output boasts the schema $(D_1, D_2, D_3,val)$,
where $D_1$ takes matching dimension from $RID_A$, $D_2$ takes relevant dimension from $CID_A$, $D_3$ takes relevant dimension from $CID_B$, and $val$ is the evaluation of the merge function on the matching entries from $\mA$ and $\mB$. The join output may degrade to a normal matrix when $\gamma$ contains two or more predicates on the join dimensions.





\subsection{Cross-product on Matrices}
The cross-product operator is a special join operator between two matrices, where the join predicate $\gamma$ is empty, i.e.,
$\mA \kron \mB = \mA \Join_f \mB$.
If both input matrices are viewed as relations, 
this operator is essentially the Cartesian product.
The cross-product is widely used in tensor decomposition and applications, e.g., Kronecker product~\cite{Kolda:tensor}.
\system's optimizer infers the output schema from the join predicate, which is empty for cross-product. Therefore, a cross-product produces a 4th-order tensor from two input matrices. 
For example, $C(D_1, D_2, D_3, D_4, val)$ is the schema of $\tensor C = \mA \kron \mB$.
The first two dimensions $(D_1, D_2)$ inherit from $\mA$, and the last two dimensions $(D_3, D_4)$ inherit from $\mB$.

The execution of a cross-product is conducted in a fully parallel manner. \system~stores a matrix in blocks of equal size, where the blocks are distributed among all the workers in a cluster. First, each block of the smaller matrix  is duplicated $p$ times, where $p$ is the number of partitions from the larger matrix. Next, each duplicated block is sent to a corresponding partition of the larger matrix. Finally, each worker performs join operation locally without further communication. The 4th-order tensor is stored as a series of block matrices in a distributed manner. The detailed physical storage is described in Section~\ref{physical-storage-matrix}.

\subsection{Join on Two Dimensions}
\begin{figure}
\centering
\includegraphics[width=0.6\columnwidth]{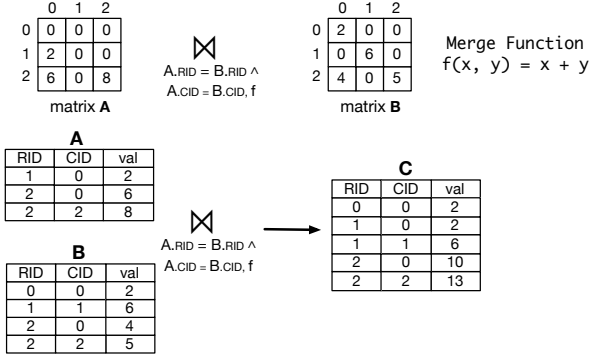}
\caption{Direct overlay of two sparse matrices.}
\vspace{-1em}
\label{fig:direct_overlay}
\end{figure}

The dimension-based predicates for join on two dimensions can take two different formats:
\[
\mA \Join_{RID_A = RID_B \land CID_A = CID_B,f} \mB,
~\text{or}~ 
\mA \Join_{RID_A = CID_B \land CID_A = RID_B,f} \mB.
\]
These two predicates are the only valid formats, as a propositional formula takes dimensions from both inputs.
The former predicate can be viewed as an overlay on the two input matrices directly~(\textit{direct overlay}), while the latter can be viewed as an overlay on Matrix $\mA$ and the transpose of Matrix $\mB$~(\textit{transpose overlay}). Figure~\ref{fig:direct_overlay} illustrates an example of a  direct overlay, which can be viewed as a full-outer join in relational algebra. Conceptually, two input matrices can be regarded as two relations with the schema introduced in Section~\ref{section:schema}. The output of a join on two dimensions is a normal matrix, since two common dimensions are shared with the inputs. 

To efficiently evaluate joins on two dimensions, \system~adopts the hash join strategy to partition the matrix blocks from the inputs. By hashing the matched matrix blocks to the same worker, a worker conducts local join execution without further communications. For direct overlay, \system's query planner partitions the two input matrices using the same partitioner. For transpose overlay, the planner makes sure the partitioning scheme on one input matches the partitioning scheme on the transpose of the other.

\subsection{Join on a Single Dimension}
The dimension-to-dimension~(D2D) predicates can take four different formats:
\[
\mA \Join_{ID_A = ID_B, f} \mB,
\]
where $ID_{A/B} \in \{RID_{A/B}, CID_{A/B}\}$.
Therefore, the join output is a 3rd-order tensor, and takes the schema $(D_1, D_2, D_3, val)$.
Each entry from a matched row/column from $\mA$ is joined with the entries from the corresponding row/column from $\mB$.

\system~leverages a hash-based data partitioner to map matched row or column matrix blocks to the same worker. Suppose both input matrices are partitioned into $\ell$-by-$\ell$ square blocks. Let us consider the D2D predicate to be $``RID_A=RID_B"$. Two matrix blocks $\mA_{mp}$ and $\mB_{mq}$ are mapped to the same worker. For the $t$-th row from $\mA_{mp}$, each entry is joined with the same row from $\mB_{mq}$. Thus, a $1$-by-$\ell$ vector is produced by joining a single entry in the $t$-th row 
from $\mA_{mp}$ and an entire row from $\mB_{mq}$. 
After the join operation, the $t$-th row from $\mA_{mp}$ leads to an $\ell$-by-$\ell$ matrix block. 
Potentially, there are totally $\ell$ such matrix blocks for the join result from two input matrix blocks.

\subsection{Join on the Entries}
The entry-based~(value-to-value, or V2V) join operation takes the following format:
\[
\mA \Join_{val_A = val_B, f} \mB.
\]
The join output is a 4th-order tensor, where the first two dimensions come from the dimensions of $\mA$, and the other two from the dimensions of $\mB$.

The execution of a V2V join is conducted in a fully parallel manner, similar to the execution of cross-product.
\system~broadcasts each block of the smaller input to each partition of the larger one.
Each worker in the cluster adopts a nested-loop join strategy to compute the join locally by taking each pair of matrix blocks. The four dimensions are copies of the dimension values from matched entries.

\subsection{Join on a Single Dimension and an Entry}
The join on a single dimension and an entry operation~(D2V, or V2D) takes the following format:
\[
\mA \Join_{ID_A = val_B, f} \mB \text{, or } \mA \Join_{val_A = ID_B, f} \mB
\]
where $ID_{A/B} \in \{RID_{A/B}, CID_{A/B}\}$.
The output is a 4th-order tensor, where the 4 dimensions derive from the entry locations of the matched rows/columns, and the matched entries from the other input. To evaluate this type of join, the matrix blocks containing the matched entries are mapped to the other matrix, where the row/column dimension matches. Each worker conducts local join computation based on the matched dimensions and entries.


\subsection{Join Optimization}
Relational joins over matrix data are more complex than other relational operations. They potentially have higher computation and communication overhead. We have identified several heuristics to mitigate the heavy memory footprint and computation burden. We leverage the following features: (1) sparsity-inducing merge functions, and (2) Bloom-join for join on entries. Furthermore, we also develop a cost model to capture the communication overhead among different types of joins. By utilizing the heuristics and the cost model, \system~generates a computation- and communication-efficient execution plan for a join operation.

\textbf{Identifying Sparsity-inducing Merge Functions.} Given a general join operation, $\mA \Join_{\gamma, f} \mB$, there are two important parameters, $\gamma$ and $f$. The predicate $\gamma$ is utilized to locate the join entries, and $f$ is evaluated on the two entries for an output entry. In real-world applications, large matrices are usually sparse and structured\footnote{\url{http://www.cise.ufl.edu/research/sparse}}. For example, Figure~\ref{fig:direct_overlay} illustrates a direct overlay on two sparse matrices, where the merge function is $f(x,y) = x + y$.
We say $f(x, y)$ is a sparsity-inducing function if $f(0, \cdot) = 0$ or $f(\cdot, 0) = 0$. Thus, the summation merge function is \textit{not} sparsity-inducing since $C_{00} = A_{00} + B_{00} = 0+2=2$, which does not preserve sparsity from the left- or right-hand side. On the other hand, $f_1(x, y) = x*y$, is sparsity-inducing on both sides.

The benefit of utilizing sparsity-inducing function is obvious, as it avoids evaluating the function completely if one of the inputs contains all zeros.
Traditionally, a straw man execution plan takes two input matrix blocks. It at first converts the sparse matrix formats to the dense counterpart by explicitly filling in 0's, then evaluates the merge function on the matched entries. Note that a sparse matrix format only records nonzeros and their locations. Compressed sparse representations reduce memory consumption and computation overhead for matrix manipulations. However, the memory consumption grows significantly when a sparse matrix is stored in the dense format.

Thus, it is critical to identify the sparsity-inducing merge functions, and avoid converting a sparse matrix format to dense  format whenever possible. We consider a special family of merge functions: linear function and their linear combinations, i.e., 
$f(x,y) = g(x)*y + h(x)$, or $f(x,y) = g(y)*x + h(y)$, where both $g(\cdot)$ and $h(\cdot)$ are linear functions.
For any merge function in this family,
\system~adopts a sampling approach to identify the sparsity-inducing property. Given a merge function $f(x, y)$, we compute $t_1 = f(0, s_1)$, and $t_2 = f(0, s_2)$, where $s_1$ and $s_2$ are non-zero random numbers. If both $t_1$ and $t_2$ are 0, it can be easily confirmed that both $g(0)$ and $h(0)$ equal to 0. Thus, $f(x,y)$ is sparsity-inducing because of the linearity of function components. A similar sparsity-inducing test could be conducted on the $y$ value.

Once a sparsity-inducing function is identified, \system's planner orchestrates the execution of join operations by leveraging the sparse matrix computations. For example, if the merge function $f(x,y)$ boasts the sparsity-inducing property on the $x$ component, \system~only retrieves the nonzero entries from the left-hand side, and joins with entries from the right-hand side. All the computations are conducted in the sparse matrix format without converting to dense matrix format.

\textbf{Bloom-join on Entries.}
For a join predicate that involves a dimension on the matrix, it is efficient for \system~to locate the corresponding row matrix blocks or column matrix blocks. The irrelevant matrix blocks are not accessed during evaluation. However, the join operation becomes expensive when the join predicate contains a comparison on the entries. A straw man plan compares each pair of entries exhaustively, where a lot of computation is wasted on the mismatched entries.

\system~adopts a Bloom-join strategy when a join predicate contains matrix entries. Each worker computes a Bloom filter on the entries. Due to the existence of sparse matrices, a worker needs to determine whether to store 0's in the Bloom filter. Thanks to the heuristic of identifying zero-preserving merge functions, 0 values are not inserted into the Bloom filter if $f(x, y)$ is sparsity-inducing. After each matrix block creates its own Bloom filter, a worker picks an entry from one of the input matrix and consults the Bloom filter on the other input. If no match is detected from the Bloom filter, the join execution continues to the next entry; otherwise, a nested-loop join is conducted on the input matrices to generate the join output.





\textbf{Cost Model for Communications.}
\system~extends \oldsystem's~\cite{matfast} matrix data partitioner, and it naturally supports three matrix data partitioning schemes: \textit{Row}~(``$r$''), \textit{Column}~(``$c$''), and \textit{Broadcast}~(``$b$''). The Broadcast scheme is only used for sharing a matrix of low dimensions, e.g., a single vector. Different matrix data partitioning schemes lead to significantly different communication overhead for join executions. Therefore, we introduce a cost model to evaluate the communication costs of various partitioning schemes for join operations on matrix data.

We design a communication cost model based on the join predicates. For cross-product, we have
\begin{align*}
C_{comm}(\mA \kron \mB)=\begin{cases}
    \hfil 0, & \text{if } s_{A/B}=b.\\
    (N-1)\min\{|\mA|, |\mB|\}, & \text{otherwise}.
  \end{cases}
\end{align*}
where $s_A$ and $s_B$ are the partitioning scheme of input matrix $\mA$ and $\mB$, and $N$ is the number of workers in the cluster. $|\mA|$ refers to the size of the Matrix $\mA$, i.e., $|\mA| = mn$ if $\mA$ is an $m$-by-$n$ dense matrix; and it means \texttt{nnz}$(\mA)$ if $\mA$ is sparse. The communication cost is 0 if one of the inputs has been broadcast to every worker in the cluster. This only applies to the case when either $\mA$ or $\mB$ is a matrix of tiny dimensions. When $\mA$ and $\mB$ are partitioned in Row/Column scheme, each partition of the smaller matrix has to be mapped to every worker. Therefore, it incurs a communication cost of $(N-1)\min\{|\mA|, |\mB|\}$.

For direct overlay, we have the following cost function,
\begin{align*}
  C_{comm}(\mA \Join_{\gamma, f} \mB) =\begin{cases}
    \frac{N-1}{N}\min\{|\mA|, |\mB|\}, & \text{if } (s_{A},s_{B})=(r,c)\text{ or }(c, r).\\
   \hfil 0, & \text{otherwise}.
  \end{cases}
\end{align*}
It is clear that the direct overlay operation induces 0 communication overhead if one of the inputs is broadcast to all the workers in the cluster. Furthermore, there is no communication when both inputs are partitioned using the same scheme. Each worker receives matrix blocks with the same row/column block IDs from both inputs, and the join predicates are naturally satisfied. The communication overhead is only incurred when one of  operands is partitioned in Row scheme and the other is in Column scheme. To mitigate this partitioning scheme incompatibility, \system~repartitions the smaller matrix using the same partitioning scheme from the larger matrix.

For transpose overlay, we have a similar cost function,
\begin{align*}
  C_{comm}(\mA \Join_{\gamma, f} \mB) =\begin{cases}
    \frac{N-1}{N}\min\{|\mA|, |\mB|\}, & \text{if } (s_{A},s_{B})=
    (r,r) \text{ or }(c, c).\\
   \hfil 0, & \text{otherwise}.
  \end{cases}
\end{align*}
The cost model for a transpose overlay is very similar to that of the direct overlay case. The difference is that the communication overhead is introduced when two matrices are partitioned using the same scheme.

\begin{figure}
\centering
\includegraphics[width=0.65\columnwidth]{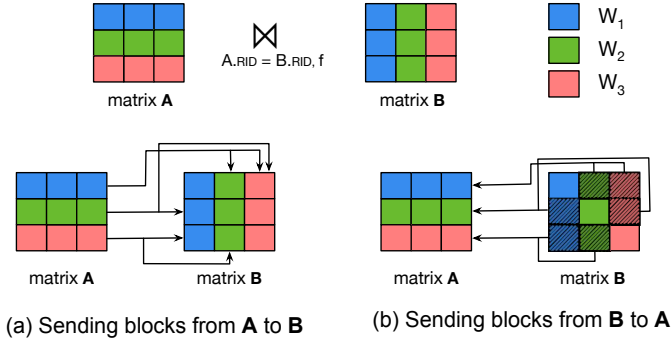}
\caption{Communication cost for D2D join.}
\vspace{-1em}
\label{fig:join_cost_1_dim}
\end{figure}

Table~\ref{tab:sinlge_dim_comm} summarizes the communication costs for joins on a single dimension. If any input matrix is broadcast to all the workers, there is no communication cost. We omit this case in  Table~\ref{tab:sinlge_dim_comm}.
Let us focus on the case when $\gamma=``RID_A=RID_B"$ and $(s_A, s_B) = (r,c)$. Figures~\ref{fig:join_cost_1_dim} illustrates that Matrix $\mA$ is partitioned in Row scheme and Matrix $\mB$ is partitioned in Column scheme. Suppose we have 3 workers in the cluster, i.e., $N=3$. Each square in the matrix is a block partition. The blocks with the same color are stored on the same worker. The join predicate requires that the entries sharing the same $RID$'s are joined together. There are two possible execution strategies based on the partitioning schemes of $\mA$ and $\mB$. Strategy I sends the blocks with the same $RID$ from $\mA$ to the workers which hold the joining blocks from $\mB$. For example, Worker $W_1$ sends all the blocks to Worker $W_2$ and $W_3$, since both $W_2$ and $W_3$ hold some joining blocks from $\mB$. Both $W_2$ and $W_3$ send the blocks in a similar manner. 
Each worker needs to send the blocks $(N\!-1)$ times to all the other workers. Thus, this strategy induces $(N\!-1)|\mA|$ communication cost.
Strategy II adopts a different policy by sending different blocks from $\mB$ to the same worker that holds the blocks with the same $RID$ from $\mA$. For example, the blocks with diagonal lines from $W_2$ and $W_3$ are sent to $W_1$ to satisfy the join predicate. As illustrated in Figure~\ref{fig:join_cost_1_dim}, all the blocks with diagonal lines introduce a communication cost of $\frac{N-1}{N}|\mB|$. Thus, the best communication cost is $\min\{(N-1)|\mA|, \frac{N-1}{N}|\mB|\}$ when $\gamma=``RID_A=RID_B"$ and $(s_A,s_B)=(r,c)$. 
The remaining entries in Table~\ref{tab:sinlge_dim_comm} can be computed with similar strategies.
Notice, the diagonal of Table~\ref{tab:sinlge_dim_comm} contains all 0's. This is because the partitioning schemes of both matrices happen to match the join predicate. 

\begin{table*}[!ht]
\centering
\scriptsize
\begin{tabular}{|c|c|c|c|c|} \hline
 & \multicolumn{4}{c|}{$(s_{A}, s_{B})$} \\  \hline
$\gamma$ & (r, r) & (r, c) & (c, r) & (c, c) \\ \hline
$RID_A=RID_B$ & 0 & $\min\{(N\!-1)|\mA|, \frac{N-1}{N}|\mB|\}$ & $\min\{\frac{N-1}{N}|\mA|, (N\!-1)|\mB|\}$ & $(N\!-1)\min\{|\mA|,  |\mB|\}$ \\ \hline 
$RID_A=CID_B$ & $\min\{(N\!-1)|\mA|, \frac{N-1}{N}|\mB|\}$ & 0 & $(N-1)\min\{|\mA|, |\mB|\}$ & $\min\{\frac{N-1}{N}|\mA|, (N\!-1)|\mB|\}$ \\ \hline
$CID_A=RID_B$ & $\min\{\frac{N-1}{N}|\mA|, (N\!-1)|\mB|\}$ & $(N-1)\min\{|\mA|,|\mB|\}$ & 0 & $\min\{(N\!-1)|\mA|, \frac{N-1}{N}|\mB|\}$ \\ \hline
$CID_A=CID_B$ & $(N-1)\min\{|\mA|,|\mB|\}$ & $\min\{\frac{N-1}{N}|\mA|, (N\!-1)|\mB|\}$ & $\min\{(N\!-1)|\mA|, \frac{N-1}{N}|\mB|\}$ & 0 \\ \hline
\end{tabular}
\captionof{table}{Communication cost of different D2D join predicates.}
\vspace{-2em}
\label{tab:sinlge_dim_comm}
\end{table*}

A join on entries has an identical cost function as a cross-product.
It is clear that the join execution incurs 0 communication cost if either $\mA$ or $\mB$ is partitioned in Broadcast scheme. For any other combination of partitioning schemes of $\mA$ and $\mB$, a worker needs to broadcast the smaller of the inputs to all the other workers. Hence, the communication cost is $(N-1)\min\{|\mA|, |\mB|\}$.

For a join on a single dimension and entries, Table~\ref{tab:sinlge_dim_val_comm} gives the communication costs for the various schemes. We use $\eta$ to denote the selectivity of entries that could match the row/column dimensions from the other matrix. Let us examine the cost function for $\gamma=``RID_A=val_B"$. 
When $(s_A, s_B) = (r,r)$, there exist two strategies to evaluate the join. Strategy I requires each worker to send its own matrix blocks of $\mA$ to the other $(N-1)$ workers, since any worker may hold a block from $\mB$ that has matched entries with the row dimension. Broadcasting $\mA$ incurs a communication cost of $(N-1)|\mA|$.
Strategy II sends the matched entries from $\mB$ to each corresponding worker that holds the blocks of $\mA$. The selectivity indicates a total amount of $\eta_B|\mB|$ of matrix data is transferred. Thus, the communication cost would be $\min\{(N-1)|\mA|, \eta_B|\mB|\}$. 
For $(s_A, s_B) = (c,r)$, the only difference is that the selected matching entries from $\mB$ are sent to all the workers in the cluster, since $\mA$ is partitioned in Column scheme. The remaining entries in Table~\ref{tab:sinlge_dim_val_comm} can be computed in a similar manner.

\begin{table*}[!ht]
\centering
\scriptsize
\begin{tabular}{|c|c|c|c|c|} \hline
 & \multicolumn{4}{c|}{$(s_{A}, s_{B})$} \\  \hline
$\gamma$ & (r, r) & (r, c) & (c, r) & (c, c) \\ \hline
$RID_A=val_B$ & $\min\{(N\!-1)|\mA|,\eta_B|\mB|\}$ & $\min\{(N\!-1)|\mA|,\eta_B|\mB|\}$ & $\min\{(N\!-1)|\mA|,N\eta_B|\mB|\}$ & $\min\{(N\!-1)|\mA|,N\eta_B|\mB|\}$ \\ \hline 
$CID_A=val_B$ & $\min\{(N\!-1)|\mA|,N\eta_B|\mB|\}$ & $\min\{(N\!-1)|\mA|,N\eta_B|\mB|\}$ & $\min\{(N\!-1)|\mA|,\eta_B|\mB|\}$ & $\min\{(N\!-1)|\mA|,\eta_B|\mB|\}$ \\ \hline
$val_A=RID_B$ & $\min\{\eta_A|\mA|, (N\!-1)|\mB|\}$ & $\min\{N\eta_A|\mA|, (N\!-1)|\mB|\}$ & $\min\{\eta_A|\mA|, (N\!-1)|\mB|\}$ & $\min\{N\eta_A|\mA|,(N\!-1)|\mB|\}$ \\ \hline
$val_A=CID_B$ & $\min\{N\eta_A|\mA|, (N\!-1)|\mB|\}$ & $\min\{\eta_A|\mA|, (N\!-1)|\mB|\}$ & $\min\{N\eta_A|\mA|, (N\!-1)|\mB|\}$ & $\min\{\eta_A|\mA|, (N\!-1)|\mB|\}$ \\ \hline
\end{tabular}
\captionof{table}{Communication cost of different D2V and V2D join predicates.}
\vspace{-2em}
\label{tab:sinlge_dim_val_comm}
\end{table*}

\textbf{Algorithm for Partitioning Scheme Assignment of Joins.}
Before we delve into the algorithm of partitioning scheme assignment for joins, we first discuss the cost function for converting distributed matrix data from one partitioning scheme to another.
Table~\ref{tab:convert_comm} illustrates the conversion costs, where $s_A$ is the partition scheme of an input, and $s'_A$ is the scheme for an output. $\xi$ denotes the case when the input data is randomly partitioned among all the workers in the cluster, e.g., round-robin. It introduces the communication cost of $|\mA|$ when re-partitioning Matrix $\mA$ from round-robin to Row/Column scheme. Broadcast scheme is more expensive and it costs $N|\mA|$, where each worker has a copy of all the matrix data.
\begin{table}[!ht]
\centering
\scriptsize
\begin{tabular}{|c|c|c|c|} \hline
 & \multicolumn{3}{c|}{$s'_{A}$} \\  \hline
$s_A$ & $r$ & $c$ & $b$ \\ \hline
$r$ & 0 & $\frac{N-1}{N}$|\mA| & $(N\!-1)|\mA|$ \\ \hline 
$c$ & $\frac{N-1}{N}|\mA|$ & 0 & $(N\!-1)|\mA|$ \\ \hline
$b$ & 0 & 0 & 0 \\ \hline
$\xi$ & $|\mA|$ & $|\mA|$ & $N|\mA|$ \\ \hline
\end{tabular}
\captionof{table}{Communication cost of converting partition schemes.}
\vspace{-2em}
\label{tab:convert_comm}
\end{table}

Given a join operation and input matrices $\mA$ and $\mB$, \system's data partitioner computes the best partition schemes for $\mA$ and $\mB$ by optimizing the following,
\begin{align*}
(s'_A,s'_B) \gets \argminl_{(s'_A,s'_B)} \{C_{comm}(\mA \Join_{\gamma, f} \mB, s'_A, s'_B)
+ C_{vt}(\mA, s_A \rightarrow s'_A) 
+ C_{vt}(\mB, s_B \rightarrow s'_B)\}.
\end{align*}
Function $C_{comm}(\mA \Join_{\gamma, f} \mB, s'_A, s'_B)$ computes the communication cost of the join according to our cost model when $\mA$ is partitioned in  Scheme $s_A$ and $\mB$ in Scheme $s_B$. Function $C_{vt}(\mA, s_A \rightarrow s'_A)$ computes the communication cost when converting $\mA$ from Scheme $s_A$ to $s'_A$. Essentially, the data partitioner adopts a grid-search strategy among all the possible combinations of partition schemes on the input matrices, and outputs the cheapest schemes for $\mA$ and $\mB$.



\section{System Implementation}
\label{sec:impl}
After 
generating 
the query execution plan and 
the
matrix data partitioning schemes,
each worker conducts matrix and relational computations locally.
\system~leverages block matrices as a basic unit for storing and manipulating matrix data in the distributed memory. We discuss briefly 
\system's
implementation on top of Spark SQL.

\subsection{Physical Storage of a Local Matrix/Tensor}
\label{physical-storage-matrix}
\begin{figure*}[!ht]
\centering
\begin{subfigure}{0.5\columnwidth}
\includegraphics[width=0.9\columnwidth]{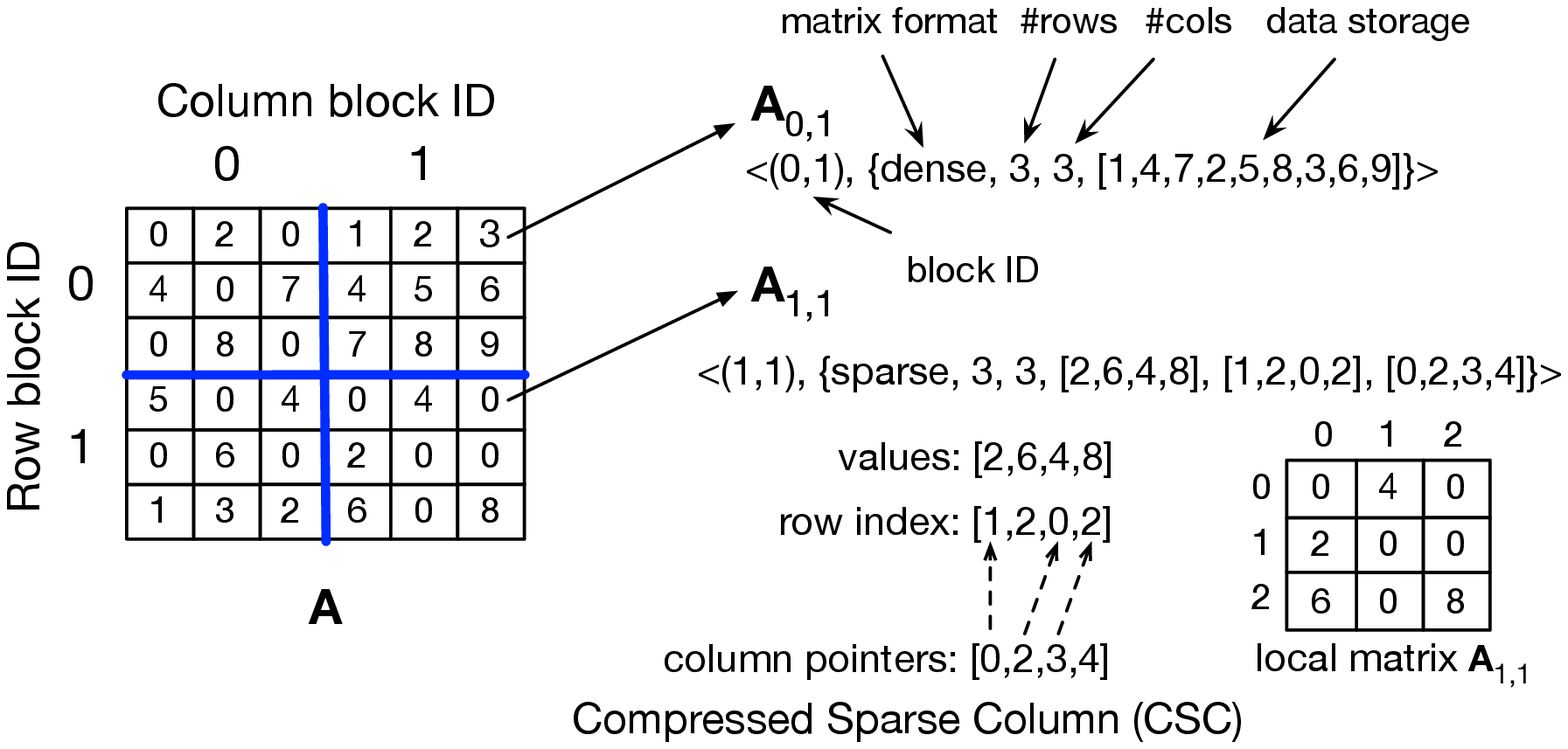}
\caption{Block matrix storage}
\label{fig:block_matrix}
\end{subfigure}
\begin{subfigure}{0.5\columnwidth}
\includegraphics[width=0.9\columnwidth]{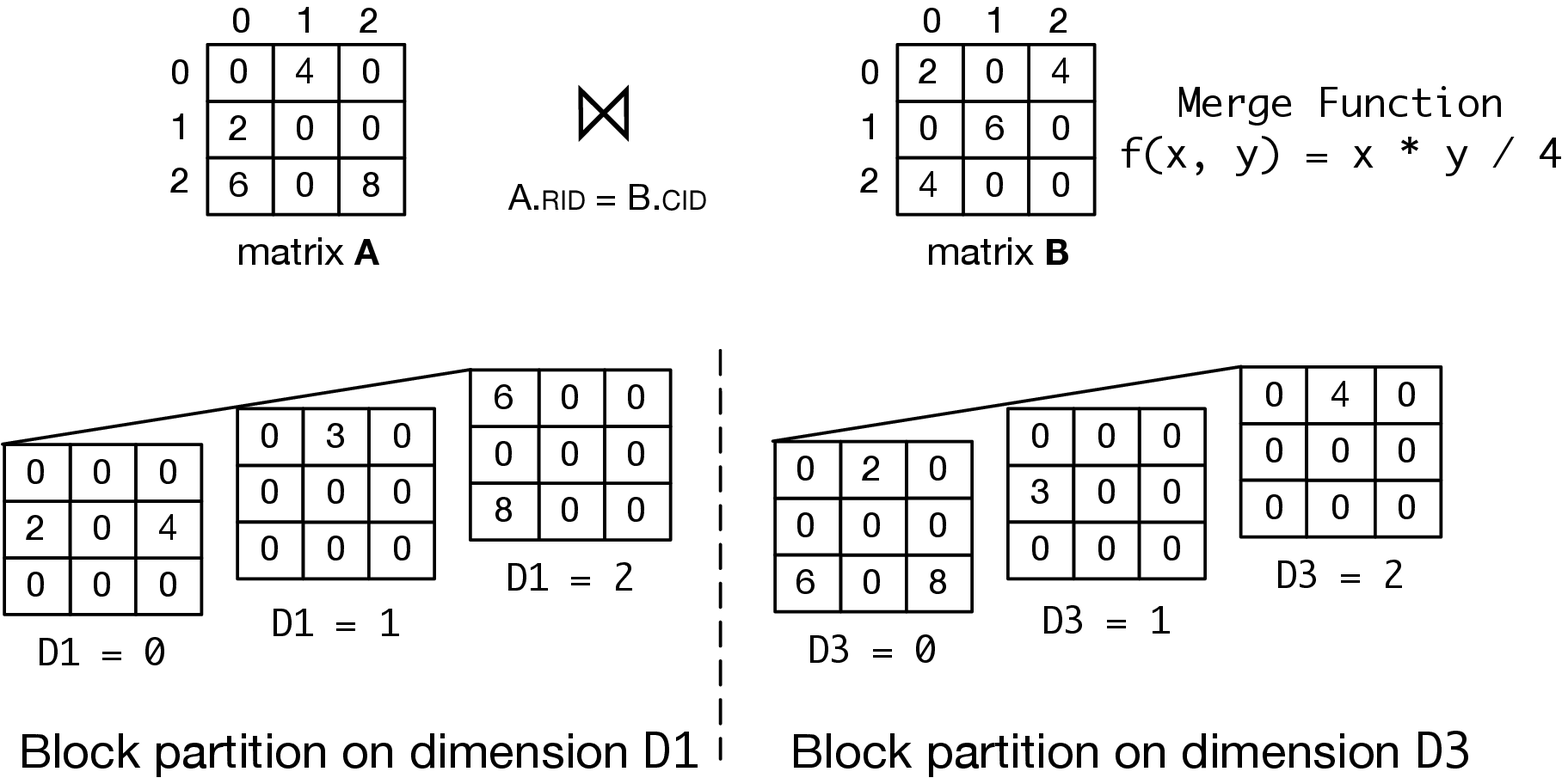}
\caption{Block tensor storage}
\label{fig:block_tensor}
\end{subfigure}
\caption{Physical storage of block matrix and tensor.}
\vspace{-2em}
\label{fig:physical_storage}
\end{figure*}

\system~partitions a matrix into smaller square blocks to store and manipulate matrix data in the distributed memory. The smaller blocks can better utilize 
the
spatial locality of 
the
nearby entries in a matrix. A matrix block is the basic unit for storage and computation. Figure~\ref{fig:block_matrix} illustrates that Matrix $\mA$ is partitioned into blocks of size $3 \times 3$, where each block is stored on a worker locally. 
For simplicity, we only consider square blocks. To fully exploit modern CPU cache size, \system~sets block size to be 1000.

Every matrix block consists of two parts; a block ID and matrix data. 
A block ID is an ordered pair, i.e., (row-block-ID, column-block-ID). The matrix data field is a quadruple, $\langle$matrix format, number of rows, number of columns, data storage$\rangle$.
A local matrix block supports both dense and sparse matrix storage formats.
For the sparse format, the nonzero entries are stored in Compressed Sparse Column~(CSC), and Compressed Sparse Row~(CSR) 
formats.
More 
detail about local matrices is
discussed in~\cite{matfast}.

A join may produce a normal matrix or a higher-dimensional tensor, e.g., a 3rd-order tensor for joins on a single dimension. \system~utilizes block matrices to manipulate higher-dimensional tensors as well. 
Given a 3rd-order tensor, the schema of the tensor is represented as \texttt{(D1, D2, D3, val)}, where \texttt{D1}, \texttt{D2}, and \texttt{D3} denote the three dimensions. 
To leverage the block matrix storage, \system~extends the block ID component to higher dimensions, e.g., (D1, D2-block-ID, D3-block-ID), where D1 records the exact dimension value, and the rest record the matrix block IDs for the other dimensions.
Figure~\ref{fig:block_tensor} illustrates the storage layout of a 3rd-order tensor in terms of matrix blocks.
There is freedom 
in
choosing which dimension of the tensor to serve as D1, D2, or D3. 
Usually there exits an aggregation  on a certain dimension following the join. A heuristic is to choose a non-aggregated dimension as D1 in a physical block storage.
It is beneficial that each worker only needs to execute the aggregation locally without further communication.
Figure~\ref{fig:block_tensor} demonstrates two possible tensor layouts on dimension D1 and D3, respectively.
If a subsequent aggregation is performed on Dimensions D2 or D3, \system~stores the join result in block matrices with respect to Dimension D1. 


\subsection{System Design and Implementation}
\system~is implemented as a library in Apache Spark. It extends Spark SQL and its Catalyst optimizer to work seamlessly with matrix/tensor data. We provide
a
Scala API for conducting relational query processing on distributed matrix data. It uses a DataFrame~\cite{sparksql} to represent distributed matrix blocks. Each row in a DataFrame has the schema \texttt{(RowBlkID, ColBlkID, Matrix)}, where both IDs 
take
a long integer type, and \texttt{Matrix} is a user-defined type that encodes the quadruple 
in Figure~\ref{fig:block_matrix}.
For logical optimization, we extend the Catalyst optimizer with all the transformation rules for optimizing the relational operations on matrix data.
Once an optimized logical plan is produced, \system~generates one or more physical plans. By leveraging the cost model for communication, the matrix data partitioner selects the partitioning schemes for input matrices that incurs minimum communication overhead. Finally, the optimized physical plan is realized by RDD's transformation operations, e.g., \texttt{map}, \texttt{flatMap}, \texttt{zipPartitions}, and \texttt{reduceByKey}, etc. The RDD partitioner class is extended with three partitioning schemes for distributed matrices, i.e., Row, Column, and Broadcast. 
Spark's fault tolerance mechanism applies naturally to \system. In case of node failure, the lost node is evicted and a standby node is chosen to recover the lost one.
An open-source version of \system~is available on GitHub\footnote{\url{https://github.com/purduedb/MatRel/tree/spark-2.1-dev}}.

\section{Experimental Results}
\label{sec:experiment}
We study the performance of the optimized execution plans by conducting different relational operations on matrix data from various ML applications. The performance is measured by the average execution time.
\subsection{Experiment Setup}
The experiments are conducted on an HP DL360G9 cluster with Intel Xeon E5-2660 realized over 6 nodes. The cluster uses Cloudera 5.9 consisting of Spark 2.1 as a computational framework and Hadoop HDFS as a distributed file system. Each node has 16 cores, 32~GB of RAM, and 400~GB of local storage. The total HDFS size is 1 Terabyte. Spark was configured with 5 executors, 16 cores/executor, 16~GB driver memory, and 24~GB executor memory.

\subsection{Comparison Across Different Platforms}
\textbf{Platforms Tested.} The platforms we evaluated are:
\begin{enumerate}[(1)]
\setlength{\itemsep}{0pt}
\item \system. This is our implementation on Spark 2.1.
All computations are written in Scala using the extended DataFrame API.
The experiments are conducted on MatRel and MatRel(w/o-opt), where the optimizations are turned on and off respectively.

\item Spark MLlib. This is the built-in Spark library, \texttt{mllib.linalg}. This is run on Spark V2.1 in the cluster mode. All computations are written in Scala.

\item SystemML. This is SystemML V0.13.1, which provides the option to run on top of Apache Spark. All computations are written in SystemML's DML programming language.

\item SciDB. This is SciDB V15.12.1. All computations are written in SciDB's AQL language that is similar to SQL. Dense matrix computations are conducted with \texttt{gemm()} API, and sparse matrix computations are with \texttt{spgemm()} API.


\end{enumerate}
\subsection{Datasets}
Our experiments are performed on both real-world and synthetic datasets. The three real-world datasets are: soc-pokec\footnote{\label{snap}\url{https://snap.stanford.edu/data/}}, cit-Patents$^{\ref{snap}}$, and LiveJournal$^{\ref{snap}}$. All these datasets are sparse matrices and their statistics are shown in Table~\ref{tab:graphs}. 
Furthermore, we generate extra smaller datasets, e.g., \texttt{u100k} and \texttt{d15k}. The first letter denotes the matrix type, i.e., $``\texttt{u}"$ for a sparse matrix with a uniform distribution of the nonzero entries, and $``\texttt{d}"$ for a dense matrix. 
The trailing number indicates the dimension of the square matrix. For example, 
\texttt{u100k} is a 100K-by-100K sparse matrix, where nonzero entries are uniformly distributed. 
\texttt{d15k} is a 15K-by-15K dense matrix. 

\smallskip
\hspace{-1.75em}
\begin{minipage}{\linewidth}
\scriptsize
\centering
\small
\begin{tabularx}{0.54\linewidth}{@{} C{0.8in} C{1.0in} C{1.0in} @{}}\toprule[1.5pt]
\bf Graph & \bf \#nodes & \bf \#edges \\\midrule
soc-pokec &  1,632,803 & 30,622,564 \\
cit-Patents & 3,774,768 & 16,518,978\\
LiveJournal & 4,847,571 & 68,993,773\\
\bottomrule[1.25pt]
\end {tabularx} \par
\captionof{table}{Statistics of the social network datasets.} 
\vspace{-1em}
\label{tab:graphs} 
\end{minipage}


\subsection{Performance Evaluation on Different Operators}
In our experiments, we conduct several representative computations. We manually kill the job if it fails to finish within 1 hour, or 3600s.

\textbf{Aggregation on a Gram Matrix.} A Gram matrix is computed as the inner-products of a set of vectors. It is commonly used in ML applications to compute the kernel functions and covariance matrices. Given a matrix $\mX$ to store the input vectors, the Gram matrix can be computed as $\mG = \mX^T \times \mX$. We consider two different aggregations on the Gram matrix, i.e., summation along the row dimension $\Gamma_{\texttt{sum}, r}(\mG)$, and trace computation $\Gamma_{\texttt{sum},d}(\mG)$. Code~\ref{lst:gram_scala} illustrates how to compute $\Gamma_{\texttt{sum},d}(\mG)$ in~\system.


\begin{figure}
        \centering
        \begin{subfigure}[b]{0.5\textwidth}
                \includegraphics[width=\columnwidth]{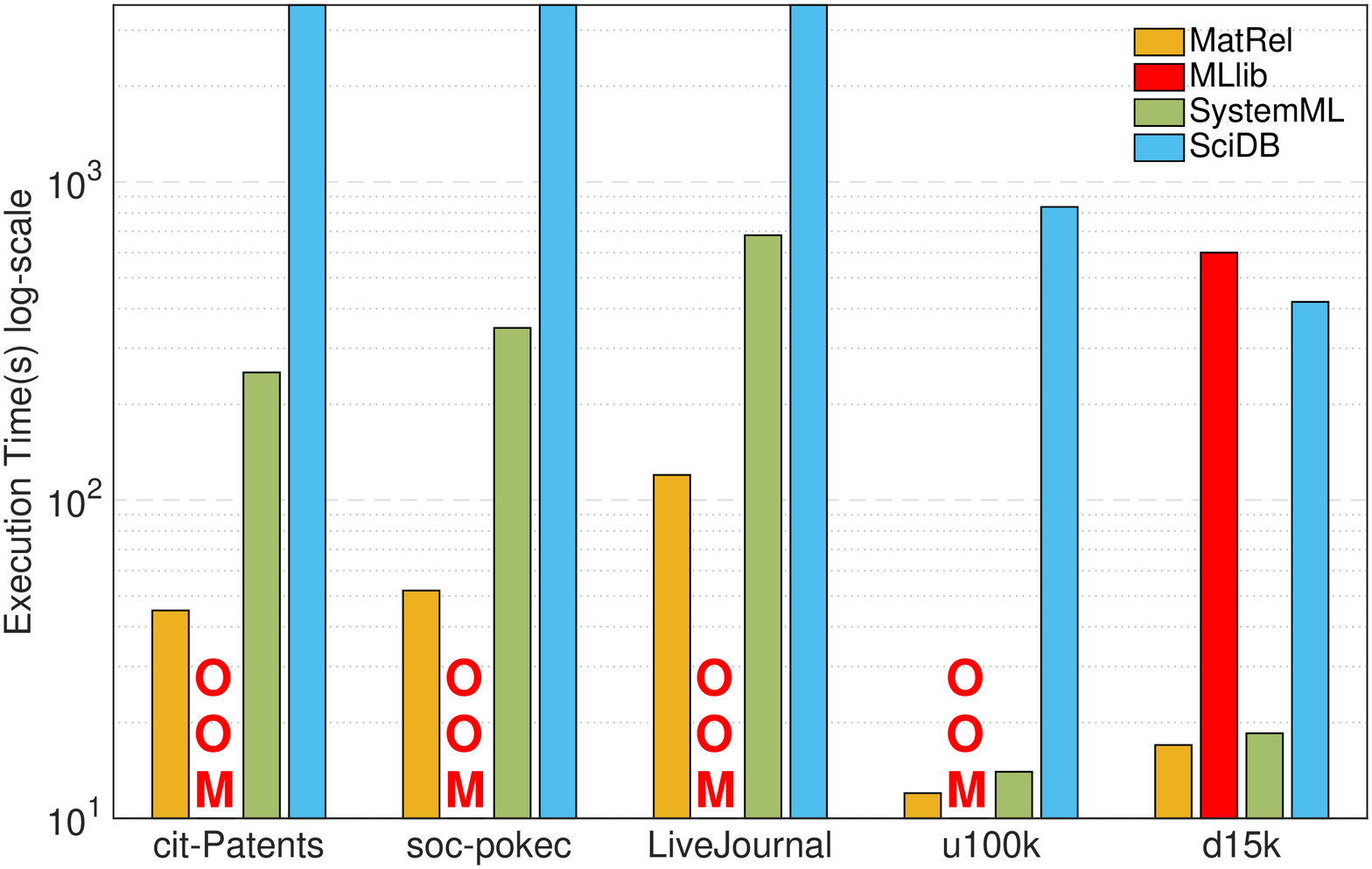}
                \caption{Execution time for $\Gamma_{\texttt{sum},r}(\mG)$}
                \label{fig:rowSum}
        \end{subfigure}%
        ~ 
        \begin{subfigure}[b]{0.5\textwidth}
                \includegraphics[width=\columnwidth]{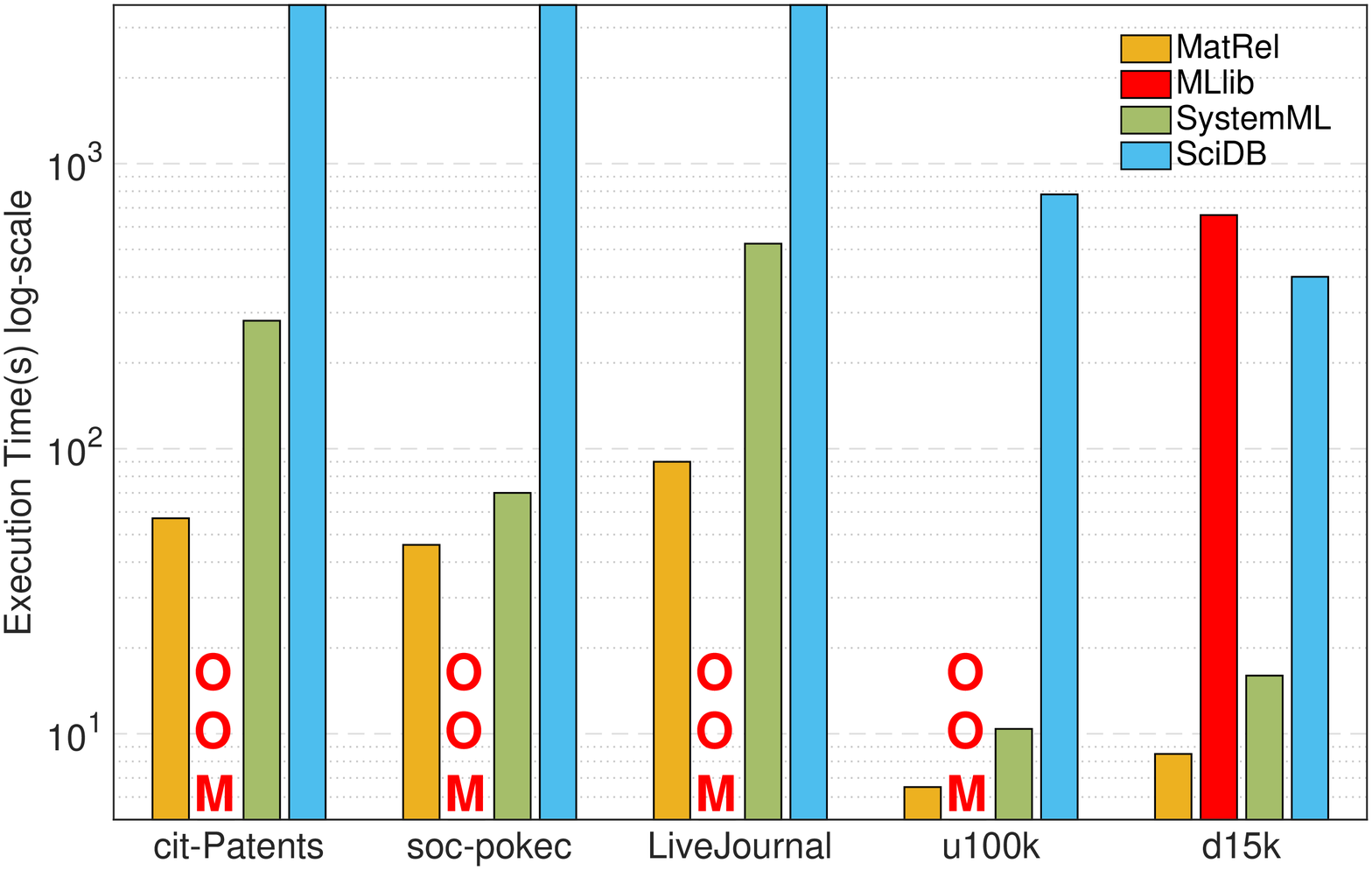}
                \caption{Execution time for $\Gamma_{\texttt{sum},d}(\mG)$}
                \label{fig:trace}
        \end{subfigure}
        \caption{Sum aggregation over matrix-matrix multiplications.}
        \vspace{-1em}
        \label{fig:agg}
\end{figure}

Figure~\ref{fig:rowSum} illustrates the execution time for different systems when performing $\Gamma_{\texttt{sum}, r}(\mX^T \times \mX)$. For all the sparse matrices, Spark MLlib throws out-of-memory~(OOM) exceptions due to the fact that MLlib converts a sparse matrix to the dense format for matrix-matrix multiplications. Even for the smallest 100K-by-100K sparse matrix with the sparsity of $10^{-5}$, it requires about 160~GB memory to store the dense matrices, which exceeds the hardware limit of our cluster.
Without pushing the aggregation below the matrix-matrix multiplications, 
SciDB runs over 1 hour to compute the product of two matrices before the aggregation. On the other hand, both MatRel and SystemML adopt the similar rewrite rule to push \texttt{sum} aggregation below the matrix-matrix multiplications. MatRel and SystemML spend 120s and 680s for the computation on LiveJournal dataset, respectively. The extra performance gain for MatRel comes from the fact that MatRel is implemented on Spark Dataset API while SystemML is run on RDDs directly. A Dataset makes extra effort for efficient data compression, serialization, and de-serialization. 
For \texttt{u100k} dataset, 
SciDB spends about 800s to compute the sparse matrix multiplication, since SciDB is not optimized for sparse matrix computations. For \texttt{d15k} dataset, all systems  finish the job within the time budget.

Figure~\ref{fig:trace} shows the execution time for various systems when performing $\Gamma_{\texttt{sum}, d}(\mX^T\times \mX)$. 
The trace computation leverages the rule to rewrite matrix-matrix multiplications in terms of matrix element-wise multiplications.
MatRel's optimizer detects the two inputs are table scans on the same matrix after query rewrite. It generates the code to compute the matrix element-wise multiplication on a single matrix without duplicate table scans. For the LiveJournal dataset, it takes 90s for MatRel and 525s for SystemML to complete the query execution. Other systems spend similar amount of time as the previous query since they fail to leverage the more efficient query rewrite rule. 

\textbf{Selection over Matrix Operations.} Least squares linear regression~(LR) is a popular ML model for classification~\cite{mlbook}. The input is a feature matrix $\mX$ and a label vector $\vy$. Each label $y_i$ can be viewed as a linear combination of the feature vector $\vx_i^T$, i.e.,  $y_i \approx \vx_i^T \times \vb + \varepsilon_i$, where $\vb$ is the vector of regression coefficients, and $\varepsilon_i$ is the error term.
The most common estimator for $\vb$ is the least squares estimator $\hat{\vb} = (\mX^T \times \mX)^{-1} \times \mX^T \times \vy$. Code~\ref{lst:regression_scala} demonstrates how to conduct selection on a Gram matrix and least squares linear regression in~\system.


\begin{figure}
        \centering
        \begin{subfigure}[t]{0.5\textwidth}
                \includegraphics[width=\columnwidth]{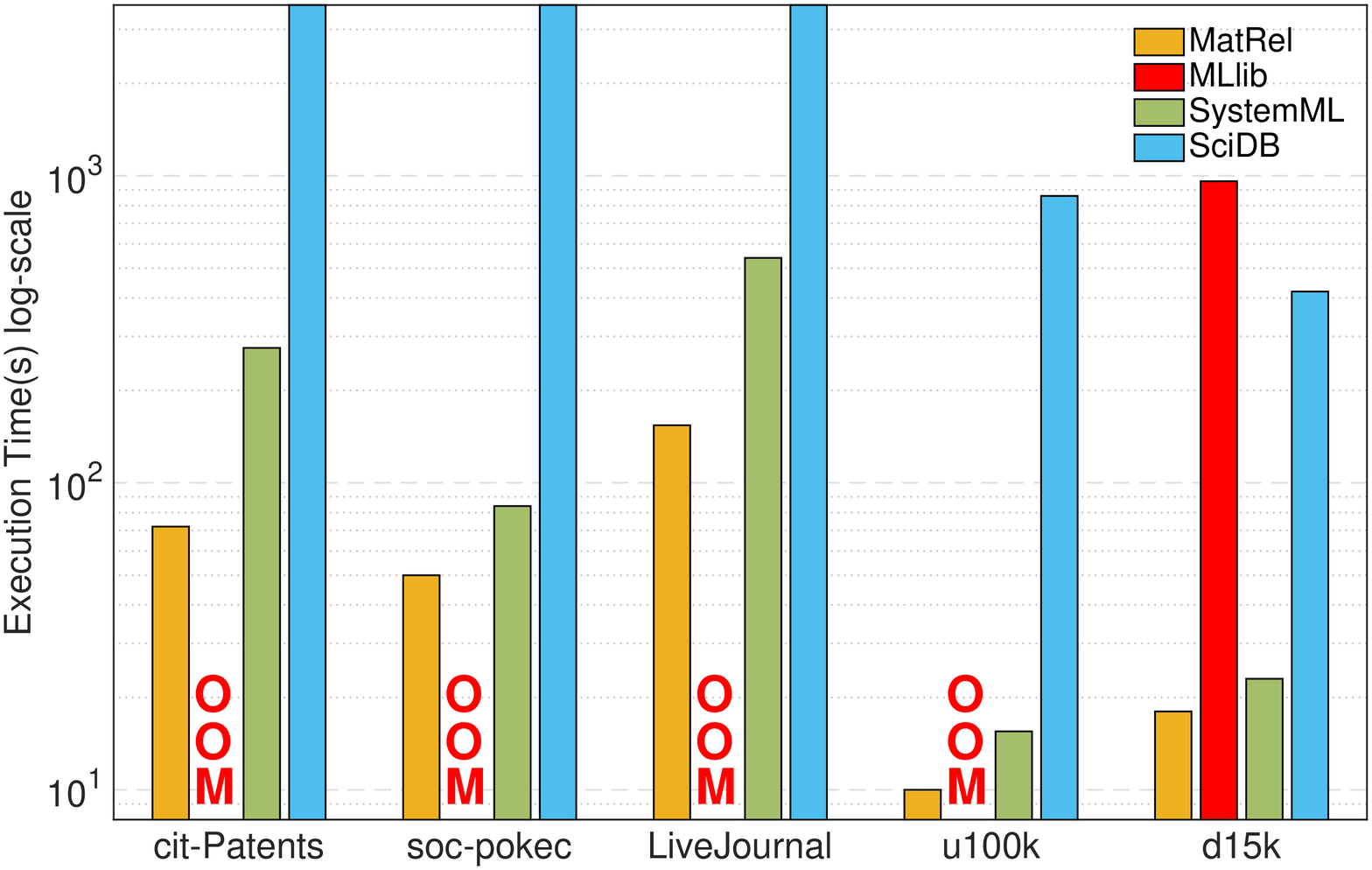}
                \caption{Execution time for LR}
                \label{fig:lr}
        \end{subfigure}%
        ~ 
        \begin{subfigure}[t]{0.5\textwidth}
                \includegraphics[width=\columnwidth]{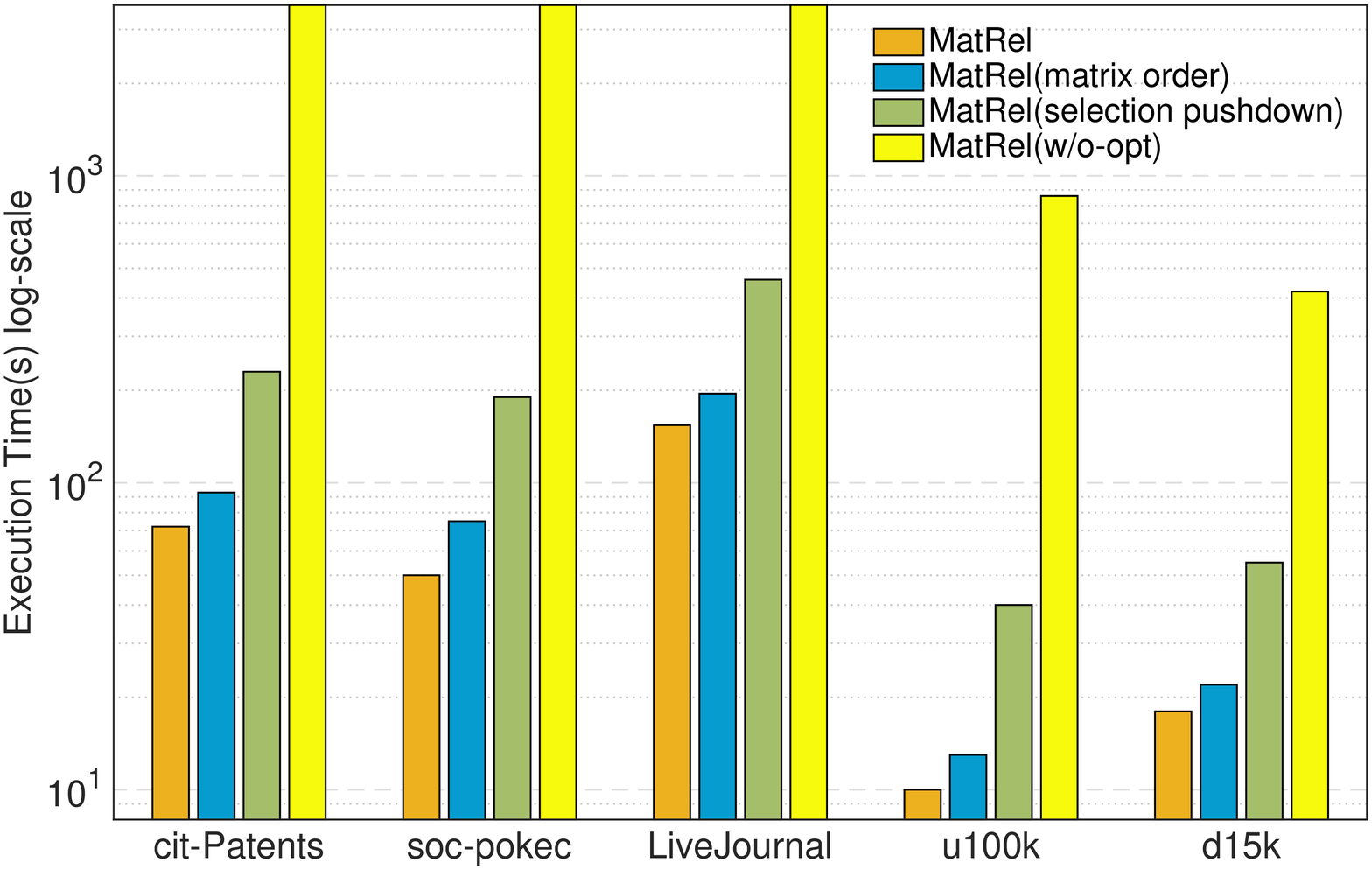}
                \caption{Effects of various optimizations in MatRel}
                \label{fig:lr_details}
        \end{subfigure}
        \caption{Selecting a row from linear regression.}
        \vspace{-1em}
        \label{fig:projection}
\end{figure}

\begin{figure}
        \centering
        \begin{subfigure}[t]{0.5\textwidth}
             \includegraphics[width=\columnwidth]{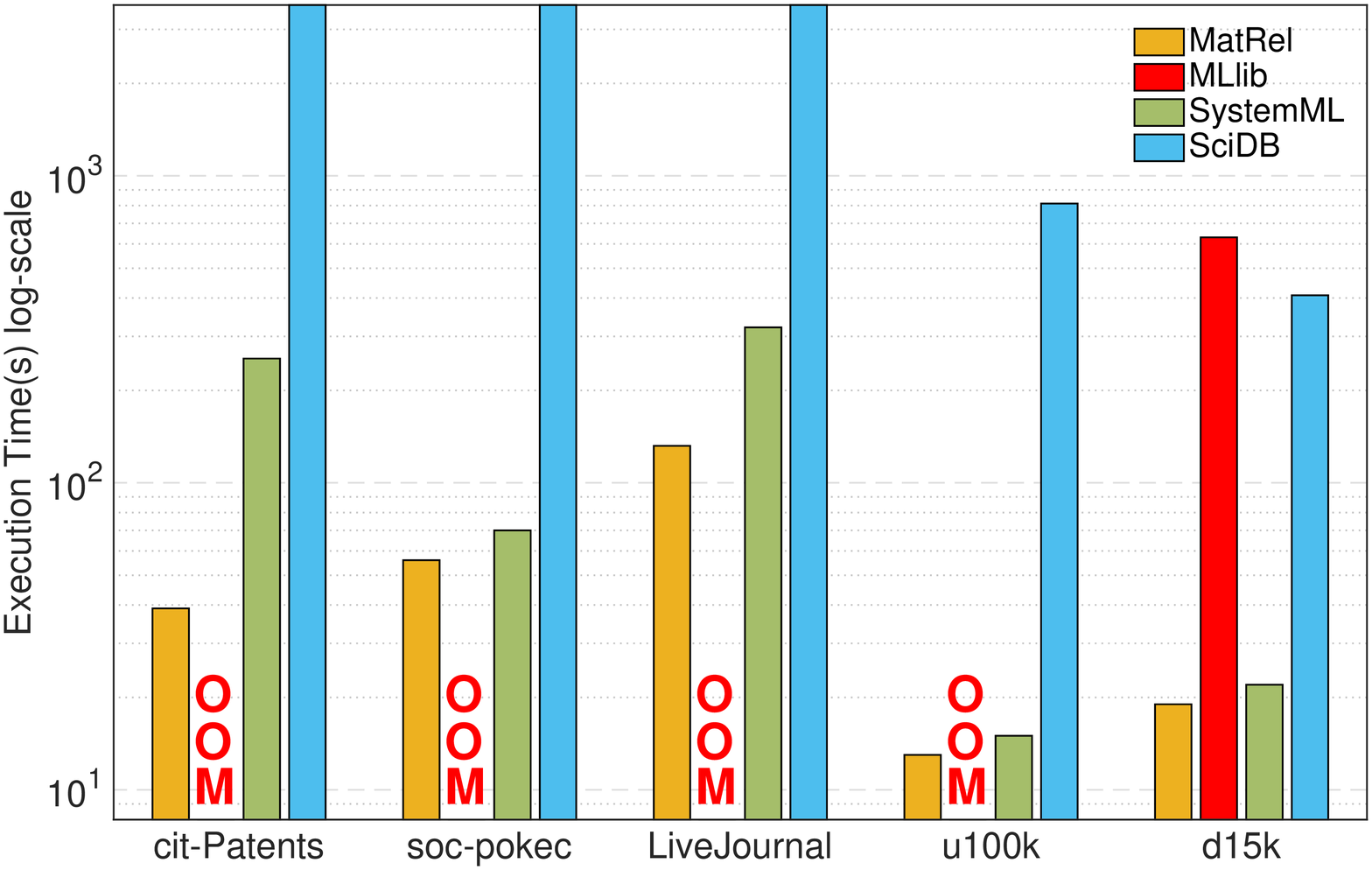}
                \caption{$\sigma_{RID=1 \land CID=1}(\mG)$}
                \label{fig:select_gram}
        \end{subfigure}%
        ~ 
        \begin{subfigure}[t]{0.5\textwidth}
                \includegraphics[width=\columnwidth]{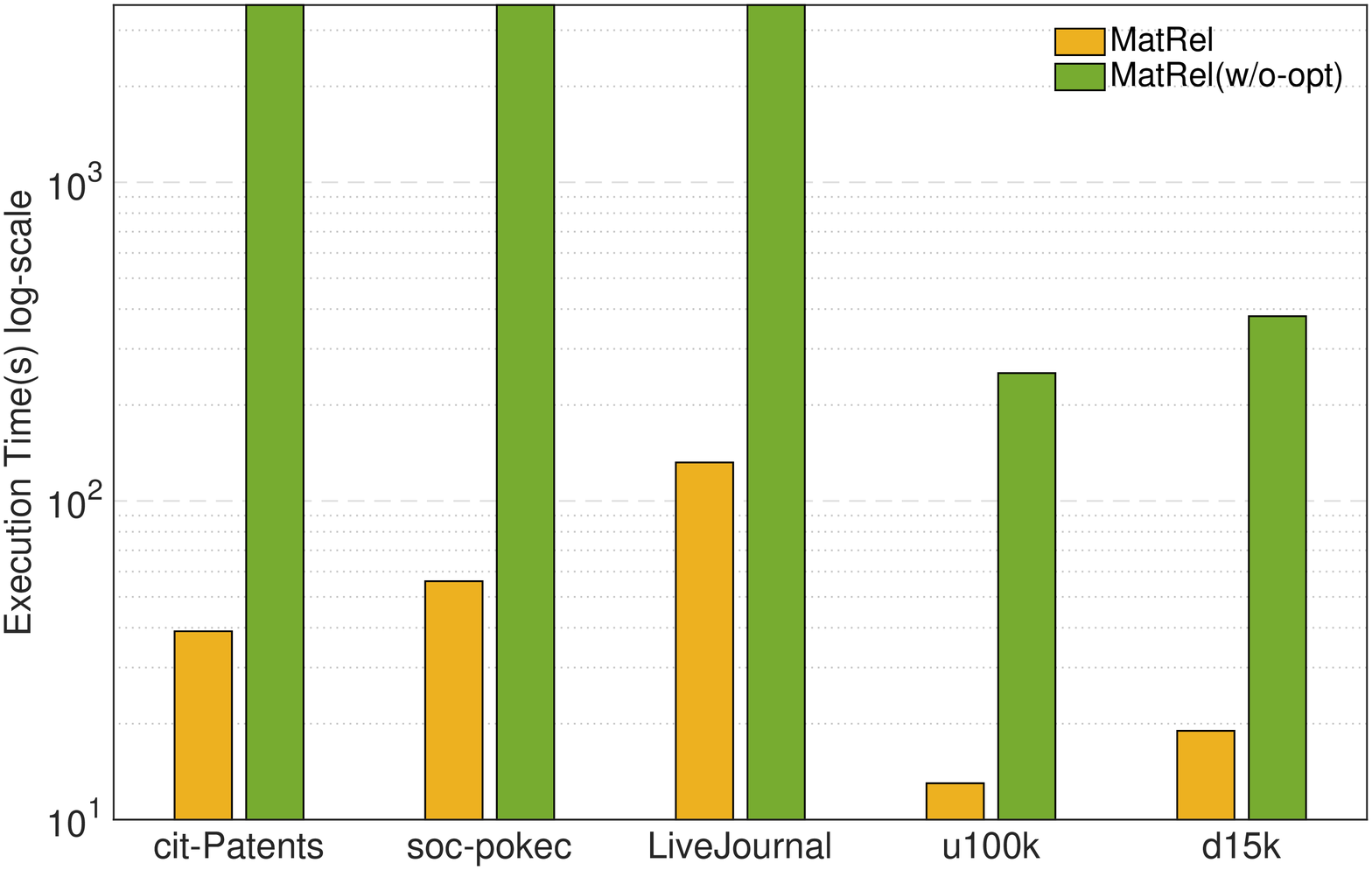}
                \caption{Pushing selection below matrix multiplications}
                \label{fig:select_gram_comp}
        \end{subfigure}
        \caption{Selecting an entry from a Gram matrix.}
        \vspace{-1em}
        \label{fig:projection}
\end{figure}

Figure~\ref{fig:lr} illustrates the execution time for selecting a row of $\hat{\vb}$. 
For an efficient evaluation of the coefficient vector $\hat{\vb}$, a good execution plan should compute $\mX^T \times \vy$ before multiplying with $(\mX^T \times \mX)^{-1}$, since multiplying $\mX^T$ with $\vy$ results in a lower dimension matrix. MatRel adopts this optimized evaluation plan. It also selects a single row on $(\mX^T \times \mX)^{-1}$. 
Spark MLlib throws OOM exceptions for all the sparse matrices due to the inefficient implementation on matrix multiplications. SystemML adopts a similar strategy for selection pushdown. 
However, the implementation on RDD prohibits SystemML from obtaining the better performance achieved by MatRel that implements the operations on Datasets. SciDB does not generate an efficient execution plan for LR computation, and performs the selection only \textit{after} the complete evaluation of $\hat{\vb}$. Figure~\ref{fig:lr_details} gives the different effects of optimizations that MatRel has adopted. Especially, we manually turn off the optimizations of order selection on matrix chain multiplications, and relational selection pushdown below matrix multiplications, e.g., MatRel~(matrix order) means turning on the optimization of order selection on matrix chain multiplications only.

Figure~\ref{fig:select_gram} gives the execution time for evaluating a selection on a matrix entry of the Gram matrix.
All the systems conduct Gram matrix computation in the same manner. With selection pushdown, MatRel and SystemML are able to finish the computation in 39s and 254s for the cit-Patents dataset. When selecting $(i, j)$-th entry on $\mX^T \times \mX$, MatRel's optimizer rewrites the query plan to $(\sigma_{CID=i}(\mX))^T \times \sigma_{CID=j}(\mX)$. The rewritten query plan avoids evaluation of a matrix transpose by a vector transpose.
SciDB cannot finish the evaluation of matrix multiplication $\mX^T \times \mX$ in 3600s. Figure~\ref{fig:select_gram_comp} illustrates the effect of selection pushdown below matrix multiplications on MatRel.

\textbf{Cross-product.} The Kronecker product is a generalization of outer-product from vectors to matrices. It is widely used in tensor decomposition and applications~\cite{Kolda:tensor}. Formally, given an $m$-by-$n$ matrix $\mA$ and a $p$-by-$q$ matrix $\mB$, the Kronecker product $\mA \kron \mB$ is the $mp$-by-$nq$ block matrix, 
\[
\mA \kron \mB = 
\begin{bmatrix} 
    a_{11}*\mB & \dots & a_{1n}*\mB \\
    \vdots & \ddots & \vdots \\
    a_{m1}*\mB& \dots & a_{mn}*\mB 
\end{bmatrix}.
\]
The Kronecker product is \textit{essentially} the cross-product between matrix $\mA$ and $\mB$ with the merge function of $f(x,y) = x*y$.
We compare MatRel and SciDB for the performance of the Kronecker product computation, since all the other systems do not support joins. Code~\ref{lst:kronecker_scala} illustrates how to compute Kronecker product in~\system.
The Kronecker product is an expensive operation that consumes lots of resources. 
Therefore, we generate another 25K-by-25K sparse matrix \texttt{u25k}, where nonzero entries are uniformly distributed with sparsity of $10^{-6}$. Table~\ref{tab:cross_product} demonstrates the execution time for different systems on various cross-product tasks. 
Both MatRel(w/o-opt) and SciDB first conduct the cross-product computation on each pair of matching entries. Then, they evaluate the merge function on the matched entries. On the other hand, MatRel's optimizer identifies that merge function $f(x,y)$ has the zero-preserving property. Thus, MatRel only computes the cross-product using the nonzero entries from the sparse input. MatRel spends about 294s to finish the evaluation of the Kronecker product between a dense matrix and a sparse one, while MatRel(w/o-opt) and SciDB cannot finish the job within 1 hour. MatRel spends only 44s for computing the Kronecker product between two sparse matrices.
When both inputs are dense matrices, no optimizations can be applied, and no system can finish the job successfully. There are totally $15^{4}\times 10^{12}$ entries when computing the Kronecker product between two 15K-by-15K dense matrices, which costs about $4 \times 10^5$ TB. This is far beyond the total amount of memory of our cluster, or even the disk space. Thus, MatRel throws OOM exceptions after 5 minutes, and SciDB throws no space left on device~(NSLOD) errors after 3 hours.

\vspace{0.5em}
\hspace{-2em}
\smallskip
\begin{minipage}[!t]{\linewidth}
\centering
\small
\begin{tabularx}{0.57\linewidth}{@{} C{.80in} C{.55in} C{.85in} C{.55in} @{}}\toprule[1.5pt]
\bf Dataset & \bf MatRel & \bf MatRel(w/o) & \bf SciDB \\\midrule
\texttt{u25k} $\kron$ \texttt{d15k} &  294s & > 1h & > 1h \\
\texttt{u25k} $\kron$ \texttt{u25k} & 44s & > 1h & > 1h \\
\texttt{d15k} $\kron$ \texttt{u25k} & 312s & > 1h & > 1h \\
\texttt{d15k} $\kron$ \texttt{d15k} & OOM & OOM & NSLOD \\
\bottomrule[1.25pt]
\end {tabularx} \par
\captionof{table}{The Kronecker product on different systems} 
\vspace{-1em}
\label{tab:cross_product} 
\end{minipage}

\textbf{Join on Dimensions.} Many real-world applications rely on joins on dimensions of big matrices, e.g., raster data overlay analysis. We examine two different join predicates: Equi-join on two dimensions, and equi-join on a single dimension. Among all the systems we compare with, only SciDB provides similar functionality with \texttt{join()} and \texttt{cross\_join()}.
The \texttt{join()} operator combines the entires of two input matrices at matching dimension values. The \texttt{cross\_join()} operator computes the cross-product of the two input matrices, and applies equality predicates to pairs of dimensions.

\begin{figure}
        \centering
        \begin{subfigure}[b]{0.5\textwidth}
                \includegraphics[width=\columnwidth]{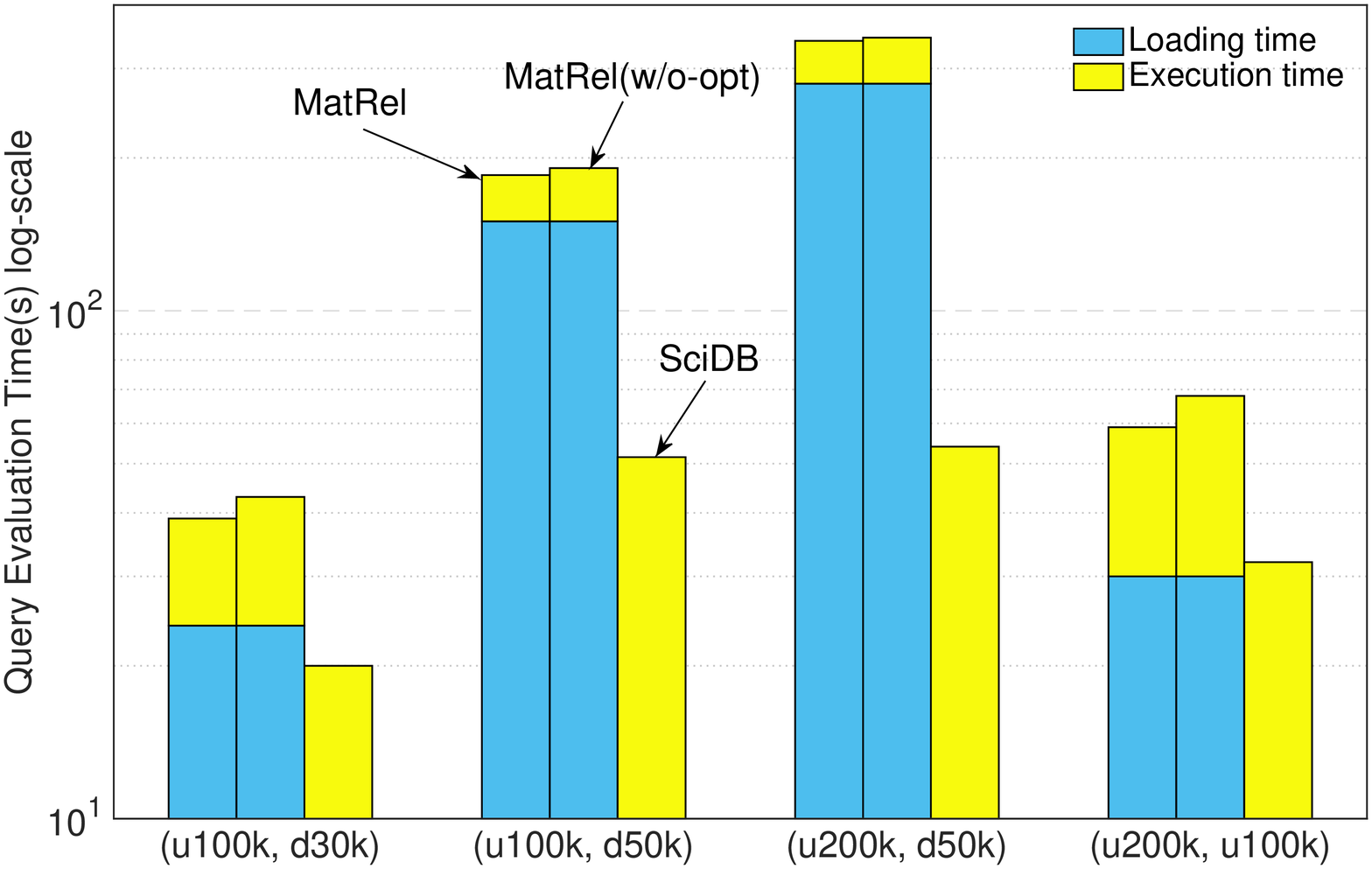}
                \caption{Direct overlay}
                \label{fig:direct}
        \end{subfigure}%
        ~ 
        \begin{subfigure}[b]{0.5\textwidth}
                \includegraphics[width=\columnwidth]{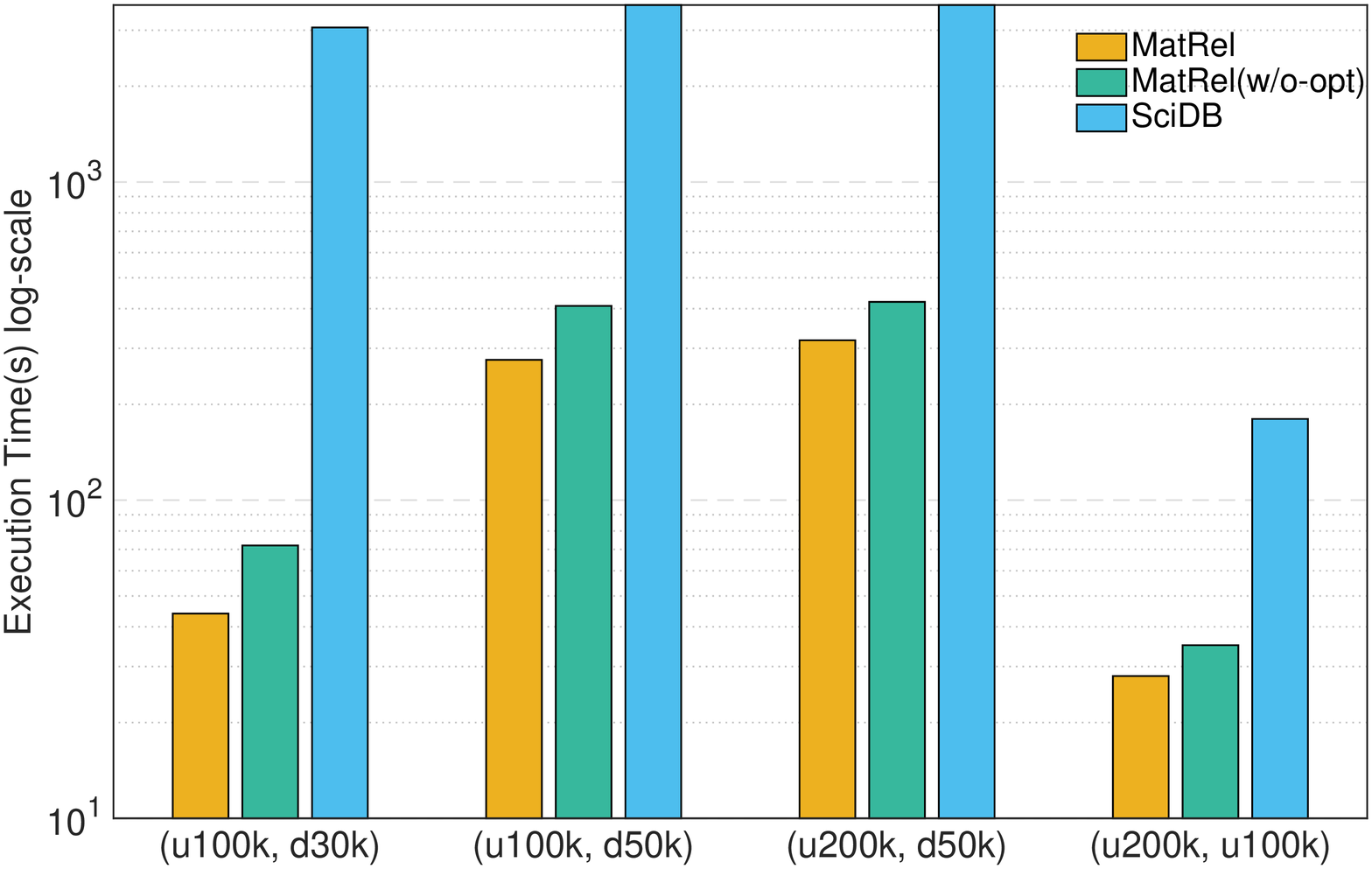}
                \caption{Transpose overlay}
                \label{fig:transpose}
        \end{subfigure}
        \caption{Execution time for join on two dimensions.}
        \vspace{-1em}
        \label{fig:projection}
\end{figure}

We conduct experiments on two different types of joins: Direct overlay and transpose overlay. The \texttt{join()} operator implements direct overlay exactly. The \texttt{cross\_join()} operator is leveraged to evaluate the transpose overlay. Code~\ref{lst:join_scala} demonstrates how to compute direct overlay in~\system.
Figure~\ref{fig:direct} illustrates the execution time of MatRel and SciDB on various combinations of inputs. \texttt{d30k} and \texttt{d50k} are two dense square matrices with dimensions 30K and 50K, respectively. Both \texttt{u100k} and \texttt{u200k} are sparse matrices with the sparsity of $10^{-5}$.
Thanks to the Multidimensional Array Clustering~(MAC) technique, SciDB stores matrix partitions in a way that the matrix chunks are aligned along the dimensions. This data layout is preferred for evaluating a direct overlay as the join can be performed locally on the matched chunks. 
On the other hand, \system~does not make any assumptions on data locations when loading data from HDFS. \system~needs to shuffle the matrix blocks to fulfill the requirement of Row/Column partitioners for join executions. Figure~\ref{fig:direct} shows a detailed decomposition of the execution time of \system. The execution time is divided into two parts; the matrix data loading time and query execution time. For SciDB, there is only execution time  since the data is already partitioned. 
The \textit{actual} execution time of \system~is similar to that of SciDB. The \system(w/o-opt) uses the default hash partitioner of Spark that has a similar performance to that of \system~for direct overlay queries.

Figure~\ref{fig:transpose} gives the performance comparison for the transpose overlay queries. For this type of query, SciDB no longer benefits from the MAC technique because the matrix chunks have to be transferred to different workers to facilitate the query execution. For example, SciDB spends about 51min  to conduct the transpose overlay query when joining \texttt{u100k} and \texttt{d30k}. On the other hand, it takes about 44s for MatRel to finish the same query that shows a consistent performance as that of the direct overlay query. When testing on larger datasets, say \texttt{u100k} and \texttt{d50k}, SciDB spends about 172min to complete, while \system~only spends 276s.

\begin{figure}
        \centering
        \begin{subfigure}[b]{0.5\textwidth}
                \includegraphics[width=\columnwidth]{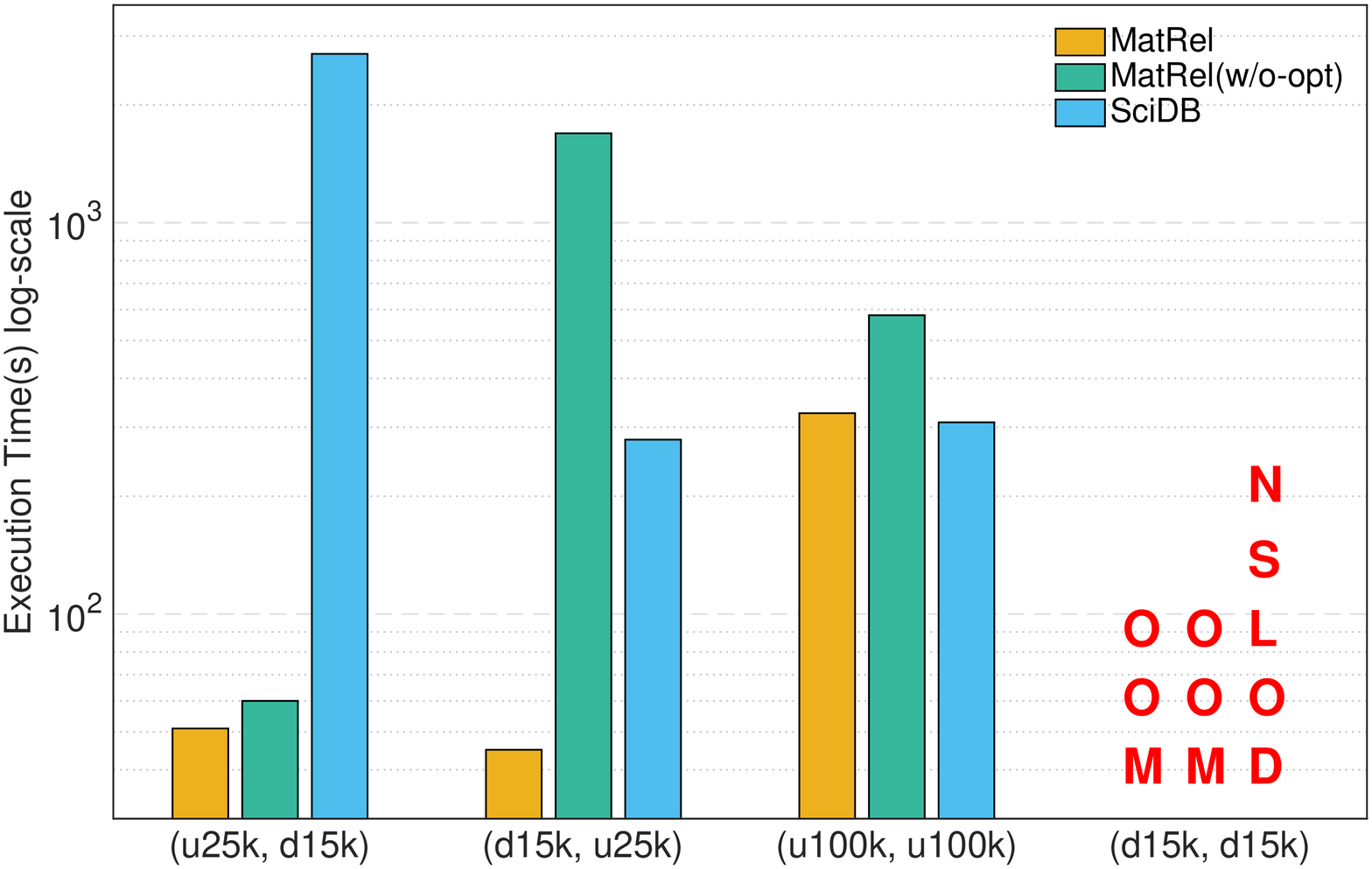}
                \caption{$\mA \Join_{RID_A = RID_B,f} \mB$}
                \label{fig:ridrid}
        \end{subfigure}%
        ~ 
        \begin{subfigure}[b]{0.5\textwidth}
                \includegraphics[width=\columnwidth]{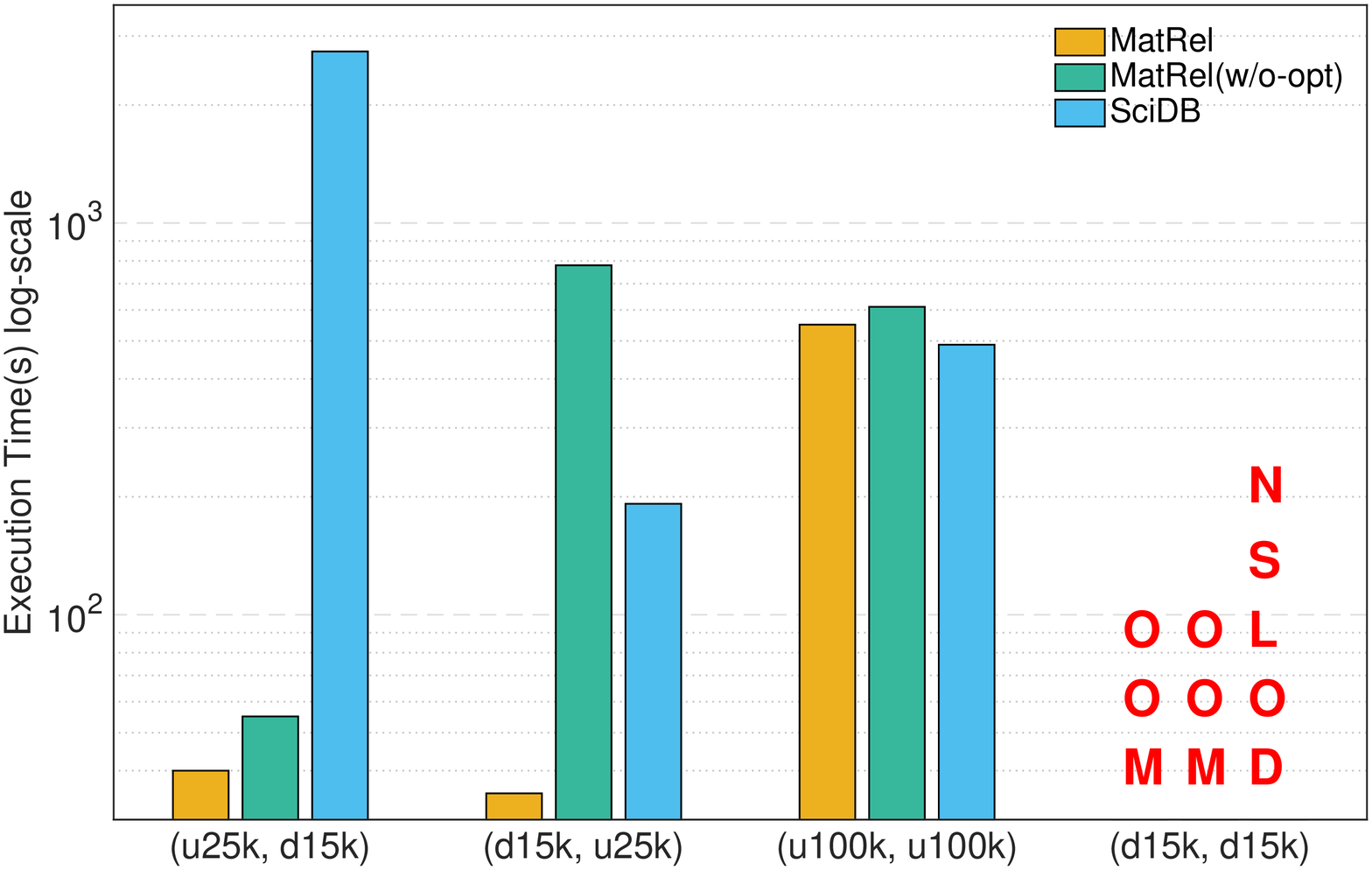}
                \caption{$\mA \Join_{CID_A = RID_B,f} \mB$}
                \label{fig:cidrid}
        \end{subfigure}

        \begin{subfigure}[b]{0.5\textwidth}
                \includegraphics[width=\columnwidth]{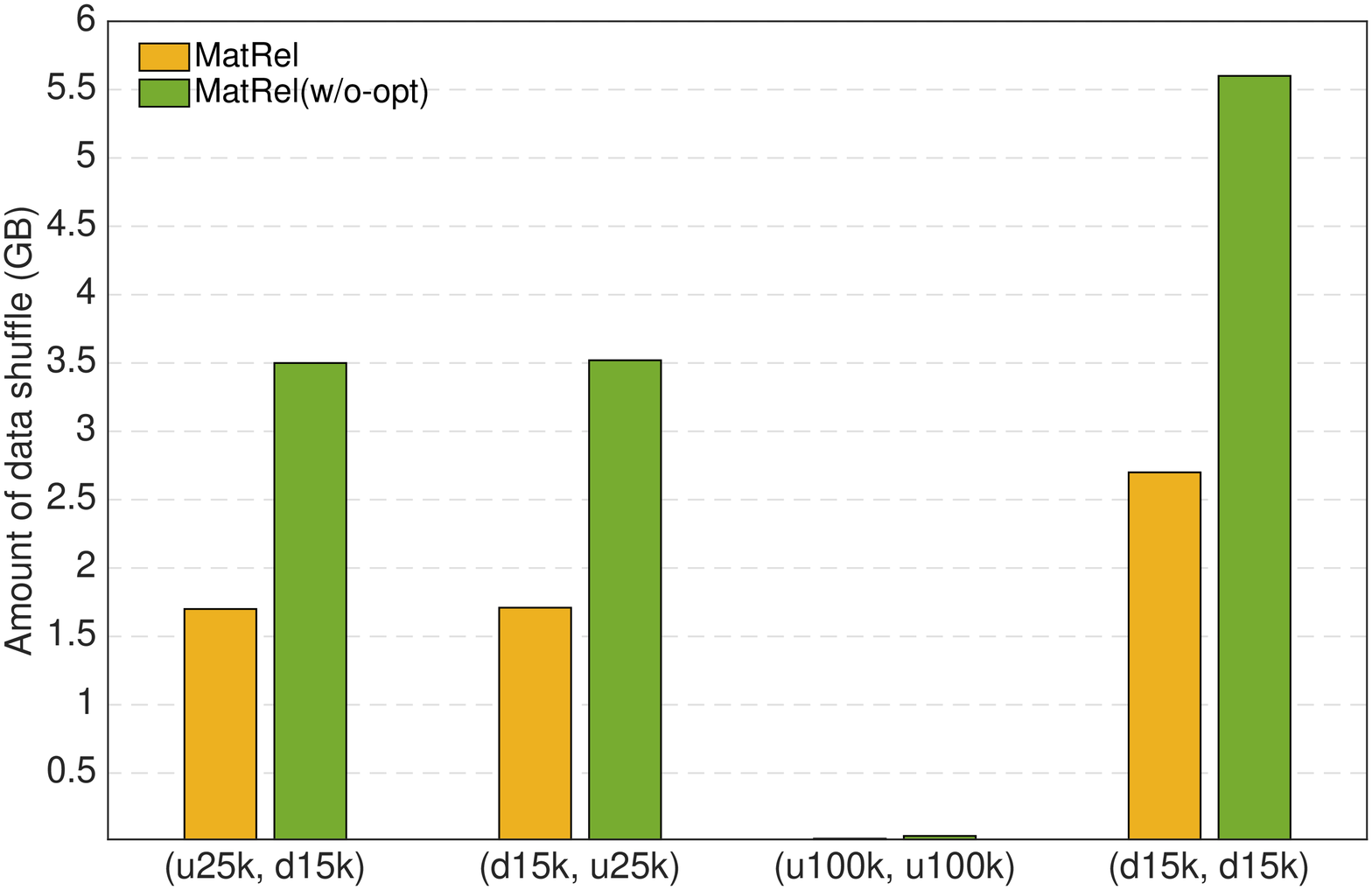}
                \caption{Data shuffle}
                \label{fig:data_shuffle}
        \end{subfigure}%
        ~ 
        \begin{subfigure}[b]{0.5\textwidth}
                \includegraphics[width=\columnwidth]{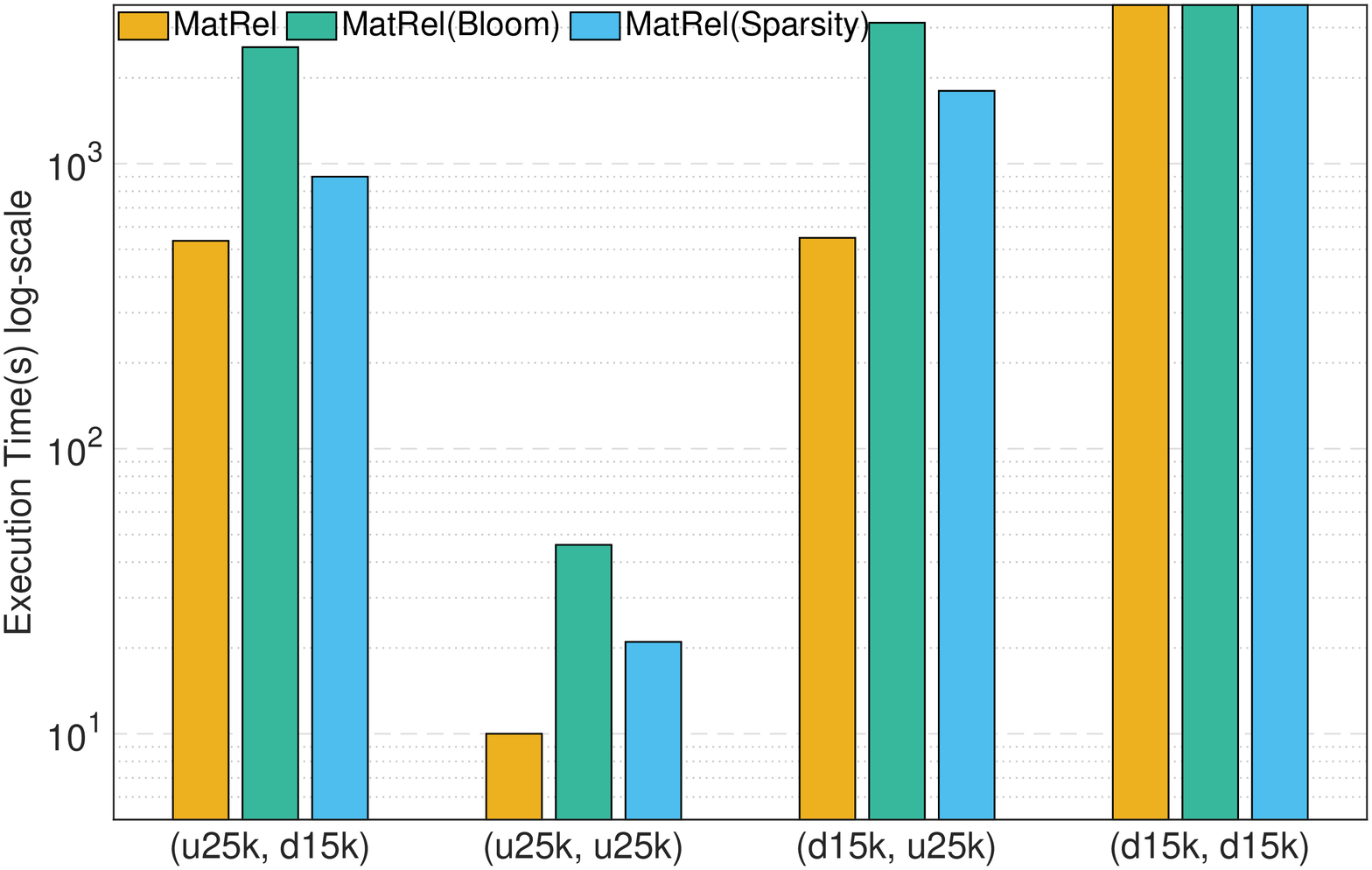}
                \caption{Join on entries}
                \label{fig:join_values}
        \end{subfigure}
        \caption{Join on a single dimension and join on entries.}
        \vspace{-2em}
        \label{fig:projection}
\end{figure}

We further perform experiments for joins on a single dimension. Figure~\ref{fig:ridrid} illustrates the performance comparison for $\mA \Join_{RID_A = RID_B, f} \mB$, where $f(x,y) = x*y$. We use different combinations of input matrices, e.g., $(\texttt{u25k}, \texttt{d15k})$. \system~leverages the sparsity-inducing property of the merge function $f(x,y)$.
\system~spends about 51s, and SciDB spends about 45min when computing the join on $\mA(\texttt{u25k})$ and $\mB(\texttt{d15k})$. \system(w/o-opt) converts a sparse matrix to the dense format, and conducts the join on the dense block, spending about 60s. SciDB's \texttt{cross\_join()} implementation is sensitive to the order of arguments, and it incurs huge overhead when the inner join matrix is a dense one. When swapping matrix $\mA$ and $\mB$, SciDB's performance improves significantly, and it only takes about 279s to complete the join. \system(w/o-opt) spends about 1690s to finish the join, since it has to duplicate the outer dense matrix multiple times and converting a sparse matrix block to a dense one for query evaluation. \system's optimizer detects both the dimensions and the sparsity of the input matrices, and shuffles the matrix with fewer nonzero entries to meet the join predicate with the other matrix for better performance.

When both inputs are sparse, \system~and SciDB exhibit similar performance. It takes longer time for \system(w/o-opt) to evaluate the query as it does not leverage the sparsity of the matrices.
When both inputs are dense matrices, no optimizations can be applied, and no system can finish the job successfully. For \texttt{d25k} dataset, each block size is 1K-by-1K and there are 225 blocks in each input. The join generates 1K new blocks for each input block. That results in about 1800~GB, which exceeds the total amount of available memory of our cluster, or even the free disk space. However, \system~throws OOM exceptions after about 5 minutes, while SciDB throws NSLOD errors after 3 hours.

Figure~\ref{fig:cidrid} illustrates the performance comparison for the join predicate of $``CID_A = RID_B"$. All the systems show similar trends as Figure~\ref{fig:ridrid}. 
Figure~\ref{fig:data_shuffle} depicts the amounts of data shuffle for \system~and \system(w/o-opt) when the join predicate is $``RID_A = RID_B"$. According to our communication cost model, \system's optimizer partitions both $\mA$ and $\mB$ using \textit{Row} scheme. \system(w/o-opt) relies on Spark's default hash partitioner to distribute the matrix among all the workers. For a load-balanced execution, we need to repartition the matrices to satisfy the join predicate. Figure~\ref{fig:data_shuffle} verifies our cost model since \system(w/o-opt) shuffles twice as much data as \system.

\textbf{Join on Entries.}
For all the systems we've compared with, only \system~supports the joins when the predicate involves matrix entries. Code~\ref{lst:join_entries_scala} shows how to compute join on entries between two matrices in~\system.
We leverage the sparsity-inducing merge function $f(x,y)=x*y$ for this query. 
\system~takes advantages of two optimizations for this kind of queries, i.e., identifying sparsity-inducing merge functions, and leveraging a Bloom-join on matrix blocks. Figure~\ref{fig:join_values} demonstrates the performance of joins on matrix entries of \system~when turning on and off the optimizations. \system(Bloom) leverages a Bloom filter for probing the entries of the inner matrix blocks without exploiting the sparsity of a sparse input. On the other hand, \system(sparsity) examines non-empty blocks in the inner matrix when conducting the join. In general, \system(sparsity) achieves a better performance than \system(Bloom) when there exists a sparse input. 
When both inputs are dense matrices, the optimization of sparsity-inducing merge function cannot apply. All the three variants of \system~cannot finish the query within 1 hour.

\textbf{Poisson Non-negative Matrix Factorization~(PNMF).} 
PNMF~\cite{nmf-www} is a popular model for dimension reduction, and it tries to approximate a sparse input matrix $\mA$ with two factor matrices $\mW$ and $\mH$ of low rank $k$, typically 10 -- 200. The computation steps are
\begin{align*}
\mW &\leftarrow \mW * ((\mA / (\mW \times \mH)) \times \mH^T) / (\mE \times \mH^T), \\
\mH &\leftarrow \mH * (\mW^T \times (\mA / (\mW \times \mH))) / (\mW^T \times \mE),
\end{align*}
where $\mA_{m \times n}$ is the input sparse matrix, $\mW_{m \times k}$ and $\mH_{k \times n}$ are two factor matrices, and $\mE$ is an $m$-by-$n$ matrix, where $E_{ij} = 1$. The updates for $\mW$ and $\mH$ continue until convergence. The objective function of PNMF is a combination of Euclidean distance and Kullback-Leibler divergence, i.e.,
\[
f(\mW, \mH) = \sum_{i,j}(\mW \times \mH)_{ij} - \sum_{i,j} (\mA * \log(\mW \times \mH))_{ij}
\]
where $\log(\mX)$ computes the logarithm of each entry in $\mX$.

The sparsity of each generated sparse matrix is $10^{-3}$. Table~\ref{tab:pnmf} shows the execution time per iteration of PNMF on different systems. \system~performs the best, followed by SystemML. MLlib has a better performance than SciDB for dataset of small sizes. However, MLlib does not scale for larger datasets,  since it does not adopts native matrix multiplications for sparse matrices.

\hspace{-2em}
\smallskip
\begin{minipage}[!t]{\linewidth}
\centering
\small
\begin{tabularx}{0.6\linewidth}{@{} C{.55in} C{.55in} C{.55in} C{.55in} C{.55in} @{}}\toprule[1.5pt]
\bf Dataset & \bf MatRel & \bf MLlib & \bf SystemML & \bf SciDB \\\midrule
\texttt{u25k} &  13s & 61s & 23s & 62s \\
\texttt{u50k} & 28s & 372s & 37s  & 260s\\
\texttt{u100k} & 52s & 2041s & 63s & 1092s\\
\texttt{u200k} & 148s & OOM & 160s & > 1h \\
\bottomrule[1.25pt]
\end {tabularx} \par
\captionof{table}{PNMF on different systems} 
\vspace{-1em}
\label{tab:pnmf} 
\end{minipage}

Notice that there are lots of opportunities when the sparsity-inducing property could be leveraged, e.g., $\mA / (\mW \times \mH)$ and $\mA * \log(\mW \times \mH)$. Matrix $\mA$ is sparse, and thus, it is unnecessary to evaluate the dense intermediate matrix of $\mW \times \mH$. \system~only computes the blocks of $\mW \times \mH$ that map to the corresponding locations of sparse blocks from $\mA$. A similar argument also applies to $\mA * \log(\mW \times \mH)$. Furthermore, \system~adopts the rule of aggregation pushdown below matrix multiplications when evaluating $\sum_{i,j}(\mW \times \mH)_{ij}$ of the objective function. \system~also interprets $\mE \times \mH^T$ into $\Gamma_{\texttt{sum}, r}(\mH)$. Therefore, \system~involves no matrix multiplications for evaluating the PNMF pipeline.

\section{Related Work}
\label{sec:related-work}
We review related work of relational and matrix query processing and optimizations on matrix data. 

\textbf{Matrix query processing on matrices:} There has been a long history of research on conducting efficient matrix computations in the high-performance computing community~(HPC). Existing libraries provide efficient matrix operators, e.g., BLAS~\cite{blas} and LAPACK~\cite{lapack} (and its distributed variant ScaLAPACK~\cite{scalapack}). However, they lack support for sparse matrices, and are prone to machine failures. 

Recently, many systems have been proposed to support efficient matrix computations using Hadoop~\cite{hadoop} and Spark~\cite{Zaharia:spark}. These systems provide an easy-to-use interface on multiple matrix operations, e.g., HAMA~\cite{hama}, Mahout~\cite{mahout}, MLI~\cite{icdm13:mli}, MadLINQ~\cite{madlinq}, Cumulon~\cite{cumulon}, SystemML~\cite{icde11:systemml,vldb14:hybrid,vldb16:matrix,systemml-spark,spoof}, DMac \cite{sigmod15:DMac}, and \oldsystem~\cite{matfast}. Each system provides various optimization strategies for reducing memory consumption, computation, and communication overheads. For example, SystemML~\cite{vldb16:matrix} adopts column encoding schemes and operations over compressed matrices to mitigate memory overhead when the original matrices are too large to fit in the available compute resources. \oldsystem~\cite{matfast} adopts a sampling-based approach to estimate the sparsity of matrix chain multiplications, and organizes distributed matrix partitions in a communication-efficient manner by leveraging matrix data dependencies. However, these systems focus on \textit{matrix-only} operations on big matrix data. None 
of these systems
provide
\textit{full-fledged} relational query processing on matrix data. \system~is built on \oldsystem, and provides optimized relational query processing on the distributed matrix data. 
MATLANG~\cite{icdt18} examines matrix manipulation from the view of expressive power of database query languages. It confirms 
that
the matrix operators provided by these systems are adequate for a wide range of applications.

\textbf{Matrix query processing on relations:}
A lot of work consider the problem of providing ML over relational data for specific ML algorithms. The assumption is 
that 
the input dataset can be viewed as a join result on multiple tables. The so-called ``factorized machine learning'' tries to learn a model based on the de-normalized tables without performing the join operation explicitly. For instance, \cite{regression-factor-join} aims at optimizing the linear regression model over factorized join tables. \cite{glm-factor} shows how to learn a generalized linear model over factorized join tables. Furthermore, \cite{join-feature-selection} discusses the cases when it is safe to conduct feature selection by avoiding key-foreign key joins to obtain features from all base tables. More recently, Morpheus~\cite{Morpheus} is proposed to conduct general linear algebra operations over factorized tables for popular ML algorithms. However, these systems rely on the  assumption that the input matrix is obtained from \textit{joins on multiple tables} that is \textit{not} always the case for real-world applications, e.g., recommender systems.

\textbf{Relational query processing on matrices:}
The idea of building a universal array DBMS on multidimensional arrays for scientific and numerical applications has been explored for a long time. One of the most notable efforts is Rasdaman~\cite{rasdaman}.

Recently, there are several systems built from scratch with native support for linear algebra, 
e.g.,
SciDB~\cite{sigmod10:scidb,arraystore,skew-join} and TensorFlow~\cite{tensorflow}. 
TensorFlow has limited support for distributed computation. The user has to manually map computation and data to each worker, since Tensorflow does not offer automatic work assignment~\cite{comparison-big-data-systems}. While Tensorflow is mainly designed for neural network models, it lacks well-defined relational operations on a matrix~(tensor).
SciDB supports the array data model, and adopts a share-nothing massively parallel processing~(MPP) architecture. SciDB provides a rich set of relational operations on 
array data. However, it treats each array operator individually without tuning for a series of array operators. 

The MADlib project~\cite{vldb12:madlib} 
supports
analytics, including linear algebra functionality, on top of a database system. MADlib shows the potential 
for
high-performance 
linear algebra on top of a relational database. 
Recent extensions on SimSQL~\cite{la-simsql} 
investigate
the possibility of making a small set of changes to SQL for enabling a distributed, relational database engine to become a high-performance platform for distributed linear algebra. However, the extension does not cover the optimization on a series of mixed relational and matrix operations. 
\system~explores the potential to optimize the query execution pipeline by pushing relational operators below matrix operators, and computes costs for 
the
different plans. \system's query optimizer also takes into account the optimized data layout of join operands for communication-efficient executions.

\section{Conclusions}
\label{sec:conclusion}
This paper presents
\system, an in-memory system that enables scalable relational query processing on big matrix data in a distributed setup. 
\system~supports common relational operations on big matrix data, e.g., relational selection, aggregation, join. \system's query optimizer leverages rule-based heuristics to rewrite a query into an equivalent execution plan with lower computation costs. 
For relational joins, \system~can leverage the sparsity-inducing property of the merge function and Bloom-join strategies for efficient executions. Furthermore, \system~adopts a cost model to generate communication-efficient matrix data partitioning schemes for input matrices on various join predicates. The experimental study on various applications demonstrates that \system~achieves up to two orders of magnitude performance gain compared to state-of-the-art systems.

\section{Acknowledgements}
Walid G. Aref acknowledges the support of the U.S. National Science Foundation under Grant Numbers III-1815796 and IIS-1910216.

\bibliographystyle{ACM-Reference-Format}
\bibliography{sigproc}

\appendix

\section{Code snippets}
\begin{lstlisting}[caption={Aggregation of a Gram matrix along diagonal in Scala},label={lst:gram_scala}, frame=bt]
val X = loadMatrix("in/X") // read matrix X
val tr = X.t().multiply(X).trace()
println(s"Trace of the Gram matrix is $tr")
\end{lstlisting}

\begin{lstlisting}[caption={Selection and linear regression in Scala},label={lst:regression_scala}, frame=bt]
val X = loadMatrix("in/X") // read matrix X
val y = loadMatrix("in/y") // read vector y
val pred = SelectPredicate.parse("RID=1 AND CID=1")
val g11 = X.t().multiple(X).select(pred)
println(s"(1, 1) entry of Gram matrix is $g11")
val b = X.t().multiply(X).inverse().multiply(X).multiply(y)
b.saveAsTextFile("out/b") // save vector b to disk
\end{lstlisting}

\begin{lstlisting}[caption={Kronecker product in Scala},label={lst:kronecker_scala}, frame=bt]
val A = loadMatrix("in/A") // read matrix A
val B = loadMatrix("in/B") // read matrix B
val func = (x: Double, y: Double) => {x * y} // merge function
val kron = A.cross_prod(B, func)
kron.saveAsTextFile("out/Kronecker") // save matrix kron to disk
\end{lstlisting}

\begin{lstlisting}[caption={Direct overlay in Scala},label={lst:join_scala}, frame=bt]
val A = loadMatrix("in/A") // read matrix A
val B = loadMatrix("in/B") // read matrix B
val join_type = JoinType.parse("RID=RID AND CID=CID") // parse the join condition
val func = (x: Double, y: Double) => {x * y} // merge function
val C = A.join(B, join_type, func)
C.saveAsTextFile("out/C") // save matrix C to disk
\end{lstlisting}

\begin{lstlisting}[caption={Join on entries in Scala},label={lst:join_entries_scala}, frame=bt]
val A = loadMatrix("in/A") // read matrix A
val B = loadMatrix("in/B") // read matrix B
val join_type = JoinType.parse("VAL=VAL") // parse the join condition
val func = (x: Double, y: Double) => {x * y} // merge function
val C = A.join(B, join_type, func)
C.saveAsTextFile("out/C") // save matrix C to disk
\end{lstlisting}
\end{document}